\documentclass[journal]{IEEEtran}
\usepackage{multicol}
\usepackage{graphicx}
\graphicspath{{Figures/}}
\usepackage{epstopdf}
\usepackage{acronym}
\usepackage{amsmath}
\usepackage{amsfonts}
\usepackage{color, colortbl}
\usepackage{subfigure}
\usepackage[english]{babel}
\usepackage{blindtext}
\usepackage{algorithm} 
\usepackage{algorithmic} 
\usepackage{multirow}
\usepackage{color}
\usepackage{subfigure}
\usepackage{balance}
\usepackage{caption}

\begin{document}

\newacro{3GPP}{third generation partnership project}
\newacro{4G}{4{th} generation}
\newacro{5G}{5{th} generation}

\newacro{Adam}{adaptive moment estimation}
\newacro{ADC}{analogue-to-digital converter}
\newacro{AED}{accumulated euclidean distance}
\newacro{AGC}{automatic gain control}
\newacro{AI}{artificial intelligence}
\newacro{AMB}{adaptive multi-band}
\newacro{AMB-SEFDM}{adaptive multi-band SEFDM}
\newacro{AN}{artificial noise}
\newacro{ANN}{artificial neural network}
\newacro{ASE}{amplified spontaneous emission}
\newacro{ASIC}{application specific integrated circuit}
\newacro{AWG}{arbitrary waveform generator}
\newacro{AWGN}{additive white Gaussian noise}
\newacro{A/D}{analog-to-digital}

\newacro{B2B}{back-to-back}
\newacro{BCF}{bandwidth compression factor}
\newacro{BCJR}{Bahl-Cocke-Jelinek-Raviv}
\newacro{BDM}{bit division multiplexing}
\newacro{BED}{block efficient detector}
\newacro{BER}{bit error rate}
\newacro{Block-SEFDM}{block-spectrally efficient frequency division multiplexing}
\newacro{BLER}{block error rate}
\newacro{BPSK}{binary phase shift keying}
\newacro{BS}{base station}
\newacro{BSS}{best solution selector}
\newacro{BU}{butterfly unit}

\newacro{CapEx}{capital expenditure}
\newacro{CA}{carrier aggregation}
\newacro{CBS}{central base station}
\newacro{CC}{component carriers}
\newacro{CCDF}{complementary cumulative distribution function}
\newacro{CCE}{control channel element}
\newacro{CCs}{component carriers}
\newacro{CD}{chromatic dispersion}
\newacro{CDF}{cumulative distribution function}
\newacro{CDI}{channel distortion information}
\newacro{CDMA}{code division multiple access}
\newacro{CI}{constructive interference}
\newacro{CIR}{carrier-to-interference power ratio}
\newacro{CMOS}{complementary metal-oxide-semiconductor}
\newacro{CNN}{convolutional neural network}
\newacro{CoMP}{coordinated multiple point}
\newacro{CO-SEFDM}{coherent optical-SEFDM}
\newacro{CP}{cyclic prefix}
\newacro{CPE}{common phase error}
\newacro{CRVD}{conventional real valued decomposition}
\newacro{CR}{cognitive radio}
\newacro{CRC}{cyclic redundancy check}
\newacro{CS}{central station}
\newacro{CSI}{channel state information}
\newacro{CSIT}{channel state information at transmitter}
\newacro{CSPR}{carrier to signal power ratio}
\newacro{CW}{continuous-wave}
\newacro{CWT}{continuous wavelet transform}
\newacro{C-RAN}{cloud-radio access networks}

\newacro{DAC}{digital-to-analogue converter}
\newacro{DBP}{digital backward propagation}
\newacro{DC}{direct current}
\newacro{DCGAN}{deep convolutional generative adversarial network}
\newacro{DCI}{downlink control information}
\newacro{DCT}{discrete cosine transform}
\newacro{DDC}{digital down-conversion}
\newacro{DDO-OFDM}{directed detection optical-OFDM}
\newacro{DDO-OFDM}{direct detection optical-OFDM}
\newacro{DDO-SEFDM}{directed detection optical-SEFDM}
\newacro{DFB}{distributed feedback}
\newacro{DFDMA}{distributed FDMA}
\newacro{DFT}{discrete Fourier transform}
\newacro{DFrFT}{discrete fractional Fourier transform}
\newacro{DL}{deep learning}
\newacro{DMA}{direct memory access}
\newacro{DMRS}{demodulation reference signal}
\newacro{DoF}{degree of freedom}
\newacro{DOFDM}{dense orthogonal frequency division multiplexing}
\newacro{DP}{dual polarization}
\newacro{DPC}{dirty paper coding}
\newacro{DSB}{double sideband}
\newacro{DSL}{digital subscriber line}
\newacro{DSP}{digital signal processors}
\newacro{DSSS}{direct sequence spread spectrum}
\newacro{DT}{decision tree}
\newacro{DVB}{digital video broadcast}
\newacro{DWDM}{dense wavelength division multiplexing}
\newacro{DWT}{discrete wavelet transform}
\newacro{D/A}{digital-to-analog}

\newacro{ECC}{error correcting codes}
\newacro{ECL}{external-cavity laser}
\newacro{ECOC}{error-correcting output codes}
\newacro{EDFA}{erbium doped fiber amplifier}
\newacro{EE}{energy efficiency}
\newacro{eMBB}{enhanced mobile broadband}
\newacro{eNB-IoT}{enhanced NB-IoT}
\newacro{EPA}{extended pedestrian A}
\newacro{EVM}{error vector magnitude}

\newacro{Fast-OFDM}{fast-orthogonal frequency division multiplexing}
\newacro{FBMC}{filter bank multicarrier }
\newacro{FCE}{full channel estimation}
\newacro{FD}{fixed detector}
\newacro{FDD}{frequency division duplexing}
\newacro{FDM}{frequency division multiplexing}
\newacro{FDMA}{frequency division multiple access}
\newacro{FE}{full expansion}
\newacro{FEC}{forward error correction}
\newacro{FEXT}{far-end crosstalk}
\newacro{FF}{flip-flop}
\newacro{FFT}{fast Fourier transform}
\newacro{FFTW}{Fastest Fourier Transform in the West}
\newacro{FHSS}{frequency-hopping spread spectrum}
\newacro{FIFO}{first in first out}
\newacro{FMCW}{frequency-modulated continuous wave}
\newacro{F-OFDM}{filtered-orthogonal frequency division multiplexing}
\newacro{FPGA}{field programmable gate array}
\newacro{FrFT}{fractional Fourier transform}
\newacro{FSD}{fixed sphere decoding}
\newacro{FSD-MNSF}{FSD-modified-non-sort-free}
\newacro{FSD-NSF}{FSD-non-sort-free}
\newacro{FSD-SF}{FSD-sort-free}
\newacro{FSK}{frequency shift keying}
\newacro{FTN}{faster than Nyquist}
\newacro{FTTB}{fiber to the building}
\newacro{FTTC}{fiber to the cabinet}
\newacro{FTTdp}{fiber to the distribution point}
\newacro{FTTH}{fiber to the home}

\newacro{GAN}{generative adversarial network}
\newacro{GB}{guard band}
\newacro{GFDM}{generalized frequency division multiplexing}
\newacro{GPU}{graphics processing unit}
\newacro{GSM}{global system for mobile communication}
\newacro{GUI}{graphical user interface}

\newacro{HARQ}{hybrid automatic repeat request}
\newacro{HC-MCM}{high compaction multi-carrier communication}
\newacro{HPA}{high power amplifier}

\newacro{IC}{integrated circuit}
\newacro{ICI}{inter carrier interference}
\newacro{ID}{iterative detection}
\newacro{IDCT}{inverse discrete cosine transform}
\newacro{IDFT}{inverse discrete Fourier transform}
\newacro{IDFrFT}{inverse discrete fractional Fourier transform}
\newacro{ID-FSD}{iterative detection-FSD}
\newacro{ID-SD}{ID-sphere decoding}
\newacro{IF}{intermediate frequency}
\newacro{IFFT}{inverse fast Fourier transform}
\newacro{IFrFT}{inverse fractional Fourier transform}
\newacro{IM}{index modulation}
\newacro{IMD}{intermodulation distortion}
\newacro{IoT}{internet of things}
\newacro{IOTA}{isotropic orthogonal transform algorithm}
\newacro{IP}{intellectual property}
\newacro{IR}{infrared}
\newacro{ISAC}{integrated sensing and communication}
\newacro{ISAR}{inverse synthetic aperture radar}
\newacro{ISC}{interference self cancellation}
\newacro{ISI}{inter symbol interference}
\newacro{ISM}{industrial, scientific and medical}
\newacro{ISTA}{iterative shrinkage and thresholding}

\newacro{KNN}{k-nearest neighbours}

\newacro{LDPC}{low density parity check}
\newacro{LFDMA}{localized FDMA}
\newacro{LLR}{log-likelihood ratio}
\newacro{LNA}{low noise amplifier}
\newacro{LO}{local oscillator}
\newacro{LOS}{line-of-sight}
\newacro{LPWAN}{low power wide area network}
\newacro{LS}{least square}
\newacro{LSTM}{long short-term memory}
\newacro{LTE}{long term evolution}
\newacro{LTE-Advanced}{long term evolution-advanced}
\newacro{LUT}{look-up table}

\newacro{MA}{multiple access}
\newacro{MAC}{media access control}
\newacro{MAMB}{mixed adaptive multi-band}
\newacro{MAMB-SEFDM}{mixed adaptive multi-band SEFDM}
\newacro{MASK}{m-ary amplitude shift keying}
\newacro{MB}{multi-band}
\newacro{MB-SEFDM}{multi-band SEFDM}
\newacro{MCM}{multi-carrier modulation}
\newacro{MC-CDMA}{multi-carrier code division multiple access}
\newacro{MCS}{modulation and coding scheme}
\newacro{MF}{matched filter}
\newacro{MIMO}{multiple input multiple output}
\newacro{ML}{maximum likelihood}
\newacro{MLSD}{maximum likelihood sequence detection}
\newacro{MMF}{multi-mode fiber}
\newacro{MMSE}{minimum mean squared error}
\newacro{mMTC}{massive machine-type communication}
\newacro{MNSF}{modified-non-sort-free}
\newacro{MOFDM}{masked-OFDM}
\newacro{MRVD}{modified real valued decomposition}
\newacro{MS}{mobile station}
\newacro{MSE}{mean squared error}
\newacro{MTC}{machine-type communication}
\newacro{MUI}{multi-user interference}
\newacro{MUSA}{multi-user shared access}
\newacro{MU-MIMO}{multi-user multiple-input multiple-output}
\newacro{MZM}{Mach-Zehnder modulator}
\newacro{M2M}{machine to machine}

\newacro{NB-IoT}{narrowband IoT}
\newacro{NB}{naive Bayesian}
\newacro{NDFF}{National Dark Fiber Facility}
\newacro{NEXT}{near-end crosstalk}
\newacro{NFV}{network function virtualization}
\newacro{NG-IoT}{next generation IoT}
\newacro{NLOS}{non-line-of-sight}
\newacro{NLSE}{nonlinear Schrödinger equation}
\newacro{NN}{neural network}
\newacro{NOFDM}{non-orthogonal frequency division multiplexing}
\newacro{NOMA}{non-orthogonal multiple access}
\newacro{NoFDMA}{non-orthogonal frequency division multiple access}
\newacro{NP}{non-polynomial}
\newacro{NR}{new radio}
\newacro{NSF}{non-sort-free}
\newacro{NWDM}{Nyquist wavelength division multiplexing }
\newacro{Nyquist-SEFDM}{Nyquist-spectrally efficient frequency division multiplexing}

\newacro{OBM-OFDM}{orthogonal band multiplexed OFDM}
\newacro{OF}{optical filter}
\newacro{OFDM}{orthogonal frequency division multiplexing}
\newacro{OFDMA}{orthogonal frequency division multiple access}
\newacro{OMA}{orthogonal multiple access}
\newacro{OpEx}{operating expenditure}
\newacro{OPM}{optical performance monitoring}
\newacro{OQAM}{offset-QAM}
\newacro{OSI}{open systems interconnection}
\newacro{OSNR}{optical signal-to-noise ratio}
\newacro{OSSB}{optical single sideband}
\newacro{OTA}{over-the-air}
\newacro{OTFS}{orthogonal time frequency space}
\newacro{Ov-FDM}{Overlapped FDM}
\newacro{O-SEFDM}{optical-spectrally efficient frequency division multiplexing}
\newacro{O-FOFDM}{optical-fast orthogonal frequency division multiplexing}
\newacro{O-OFDM}{optical-orthogonal frequency division multiplexing}
\newacro{O-CDMA}{optical-code division multiple access}

\newacro{PA}{power amplifier}
\newacro{PAPR}{peak-to-average power ratio}
\newacro{PCA}{principal component analysis}
\newacro{PCE}{partial channel estimation}
\newacro{PD}{photodiode}
\newacro{PDCCH}{physical downlink control channel}
\newacro{PDF}{probability density function}
\newacro{PDP}{power delay profile}
\newacro{PDMA}{polarisation division multiple access}
\newacro{PDM-OFDM}{polarization-division multiplexing-OFDM}
\newacro{PDM-SEFDM}{polarization-division multiplexing-SEFDM}
\newacro{PDSCH}{physical downlink shared channel}
\newacro{PE}{processing element}
\newacro{PED}{partial Euclidean distance}
\newacro{PLA}{physical layer authentication}
\newacro{PLS}{physical layer security}
\newacro{PMD}{polarization mode dispersion}
\newacro{PON}{passive optical network}
\newacro{PPM}{parts per million}
\newacro{PRB}{physical resource block}
\newacro{PSD}{power spectral density}
\newacro{PSK}{pre-shared key}
\newacro{PSS}{primary synchronization signal}
\newacro{PU}{primary user}
\newacro{PXI}{PCI extensions for instrumentation}
\newacro{P/S}{parallel-to-serial}

\newacro{QAM}{quadrature amplitude modulation}
\newacro{QKD}{quantum key distribution}
\newacro{QoS}{quality of service}
\newacro{QPSK}{quadrature phase-shift keying}
\newacro{QRNG}{quantum random number generation}

\newacro{RAUs}{remote antenna units}
\newacro{RBF}{radial basis function}
\newacro{RBW}{resolution bandwidth}
\newacro{ReLU}{rectified linear units}
\newacro{RF}{radio frequency}
\newacro{RMS}{root mean square}
\newacro{RMSE}{root mean square error}
\newacro{RMSProp}{root mean square propagation}
\newacro{RNTI}{radio network temporary identifier}
\newacro{RoF}{radio-over-fiber}
\newacro{ROM}{read only memory}
\newacro{RRC}{root raised cosine}
\newacro{RSC}{recursive systematic convolutional}
\newacro{RSSI}{received signal strength indicator}
\newacro{RTL}{register transfer level}
\newacro{RVD}{real valued decomposition}

\newacro{SB-SEFDM}{single-band SEFDM}
\newacro{ScIR}{sub-carrier to interference ratio}
\newacro{SCMA}{sparse code multiple access}
\newacro{SC-FDMA}{single carrier frequency division multiple access}
\newacro{SC-SEFDMA}{single carrier spectrally efficient frequency division multiple access}
\newacro{SD}{sphere decoding}
\newacro{SDM}{space division multiplexing}
\newacro{SDMA}{space division multiple access}
\newacro{SDN}{software-defined network}
\newacro{SDP}{semidefinite programming}
\newacro{SDR}{software-defined radio}
\newacro{SE}{spectral efficiency}
\newacro{SEFDM}{spectrally efficient frequency division multiplexing}
\newacro{SEFDMA}{spectrally efficient frequency division multiple access} 
\newacro{SF}{sort-free}
\newacro{SFCW}{stepped-frequency continuous wave}
\newacro{SGD}{stochastic gradient descent}
\newacro{SGDM}{stochastic gradient descent with momentum}
\newacro{SIC}{successive interference cancellation}
\newacro{SiGe}{silicon-germanium}
\newacro{SINR}{signal-to-interference-plus-noise ratio}
\newacro{SIR}{signal-to-interference ratio}
\newacro{SISO}{single-input single-output}
\newacro{SLM}{spatial light modulator}
\newacro{SMF}{single mode fiber}
\newacro{SNR}{signal-to-noise ratio}
\newacro{SP}{shortest-path}
\newacro{SPSC}{symbol per signal class}
\newacro{SPM}{self-phase modulation}
\newacro{SRS}{sounding reference signal}
\newacro{SSB}{single-sideband}
\newacro{SSBI}{signal-signal beat interference}
\newacro{SSFM}{split-step Fourier method}
\newacro{SSMF}{standard single mode fiber}
\newacro{STBC}{space time block coding}
\newacro{STFT}{short time Fourier transform}
\newacro{STC}{space time coding}
\newacro{STO}{symbol timing offset}
\newacro{SU}{secondary user}
\newacro{SVD}{singular value decomposition}
\newacro{SVM}{support vector machine}
\newacro{SVR}{singular value reconstruction}
\newacro{S/P}{serial-to-parallel}

\newacro{TDD}{time division duplexing}
\newacro{TDMA}{time division multiple access }
\newacro{TDM}{time division multiplexing}
\newacro{TFP}{time frequency packing}
\newacro{THP}{Tomlinson-Harashima precoding}
\newacro{TOFDM}{truncated OFDM}
\newacro{TSPSC}{training symbols per signal class}
\newacro{TSVD}{truncated singular value decomposition}
\newacro{TSVD-FSD}{truncated singular value decomposition-fixed sphere decoding}
\newacro{TTI}{transmission time interval}

\newacro{UAV}{unmanned aerial vehicle}
\newacro{UCR}{user compression ratio}
\newacro{UE}{user equipment}
\newacro{UFMC}{universal-filtered multi-carrier}
\newacro{ULA}{uniform linear array}
\newacro{UMTS}{universal mobile telecommunications service}
\newacro{URLLC}{ultra-reliable low-latency communication}
\newacro{USRP}{universal software radio peripheral}
\newacro{UWB}{ultra-wideband}

\newacro{VDSL}{very-high-bit-rate digital subscriber line}
\newacro{VDSL2}{very-high-bit-rate digital subscriber line 2}
\newacro{VHDL}{very high speed integrated circuit hardware description language}
\newacro{VLC}{visible light communication}
\newacro{VLSI}{very large scale integration}
\newacro{VOA}{variable optical attenuator}
\newacro{VP}{vector perturbation}
\newacro{VSSB-OFDM}{virtual single-sideband OFDM}
\newacro{V2V}{vehicle-to-vehicle}

\newacro{WAN}{wide area network}
\newacro{WCDMA}{wideband code division multiple access}
\newacro{WDM}{wavelength division multiplexing}
\newacro{WDP}{waveform-defined privacy}
\newacro{WDS}{waveform-defined security}
\newacro{WiFi}{wireless fidelity}
\newacro{WiGig}{Wireless Gigabit Alliance}
\newacro{WiMAX}{Worldwide interoperability for Microwave Access}
\newacro{WLAN}{wireless local area network}
\newacro{WSS}{wavelength selective switch}

\newacro{XPM}{cross-phase modulation}

\newacro{ZF}{zero forcing}
\newacro{ZP}{zero padding}


\title{Waveform-Defined Security: A Low-Cost Framework for Secure Communications}
\author{{Tongyang Xu,~\IEEEmembership{Member,~IEEE}}

\thanks{T. Xu is with the Department of Electronic and Electrical Engineering, University College London (UCL), London, WC1E 7JE, UK (e-mail: tongyang.xu.11@ucl.ac.uk). This work was supported by the Engineering and Physical Sciences Research Council "Impact Acceleration Discovery to Use" Award under Grant EP/R511638/1.
}}

\maketitle

\begin{abstract}

Communication security could be enhanced at physical layer but at the cost of complex algorithms and redundant hardware, which would render traditional physical layer security (PLS) techniques unsuitable for use with resource-constrained communication systems. This work investigates a waveform-defined security (WDS) framework, which differs fundamentally from traditional PLS techniques used in today's systems. The framework is not dependent on channel conditions such as signal power advantage and channel state information (CSI). Therefore, the framework is more reliable than channel dependent beamforming and artificial noise (AN) techniques. In addition, the framework is more than just increasing the cost of eavesdropping. By intentionally tuning waveform patterns to weaken signal feature diversity and enhance feature similarity, eavesdroppers will not be able to identify correctly signal formats. The wrong classification of signal formats would result in subsequent detection errors even when an eavesdropper uses brute-force detection techniques. To get a robust WDS framework, three impact factors, namely training data feature, oversampling factor and bandwidth compression factor (BCF) offset, are investigated. An optimal WDS waveform pattern is obtained at the end after a joint study of the three factors. To ensure a valid eavesdropping model, artificial intelligence (AI) dependent signal classifiers are designed followed by optimal performance achievable signal detectors. To show the compatibility in available communication systems, the WDS framework is successfully integrated in IEEE 802.11a with nearly no adding computational complexity. Finally, a low-cost software-defined radio (SDR) experiment is designed to verify the feasibility of the WDS framework in resource-constrained communications.


\end{abstract}


\section{Introduction}

\IEEEPARstart{S}{ecurity} in communications is a hot research topic aiming to prevent confidential information leakage. The challenges of secure communications exist in various \ac{OSI} layers as explained in work \cite{adversarial_survey_2016}. The lowest layer is physical layer, which deals with radio signal transmission and reception. Since radio signals are commonly broadcasted over the air, therefore the physical layer is more vulnerable to eavesdropping.

Channel dependent defence strategies \cite{adversarial_JASC_2018}, such as millimeter wave, beamforming, artificial noise and directional modulation are proposed to mitigate unauthorized eavesdropping. These solutions intend to degrade performance at eavesdroppers by exploiting channel environments. However, unlike perfect \ac{CSI} assumptions in theoretical simulations, imperfect or blind \ac{CSI} \cite{security_CM_CSI_2015} is a common situation in practical communications. Therefore, channel dependent \ac{PLS} solutions would be unreliable without accurate CSI. Beamforming has a potential beam leakage risk \cite{PLS_practical_CNS_mmWave_2015} due to imperfect beam shaping especially when legitimate users and eavesdroppers are spatially close. Moreover, beamforming does not work practically in long distance communications since radio beams would become wide at far field. Artificial noise \cite{security_AN_TWC_2008} is regarded as an efficient solution but it wastes extra power on noise generation.

Internet of things (IoT) security \cite{IoT_security_Proceeding_2015} is challenging since IoT connects a massive number of devices with practical constraints such as low-cost hardware, low-power consumption, limited signal processing capability and small size on-board memory. Most of existing physical layer security techniques are initially designed for sophisticated systems and are not suitable for resource-constrained IoT. In IoT applications, most of the traffic occurs in uplink channels, which is from IoT devices to a central receiver. Due to limited size, limited complexity, limited power and low data rate requirements, each IoT device is typically equipped with a single antenna. In this case, a \ac{MIMO} architecture \cite{IoT_security_MIMO_survey_2017} is not possible for IoT devices and therefore the traditional beamforming is not achievable. This is also the case for millimeter wave since sending a high frequency modulated signal would complicate each IoT device and consume more power. In addition, the accurate acquisition of legitimate instantaneous CSI at the transmitter (CSIT) \cite{adversarial_survey_2019} is not practical for resource-constrained IoT applications since frequently sending pilot symbols for channel estimation would reduce power efficiency and waste spectral resources. Furthermore, since eavesdroppers are normally external to IoT networks and would passively intercept signals \cite{IoT_security_Proceeding_2015}, the location and CSIT of eavesdroppers are hardly to know by the transmitter. With the development of \ac{AI}, deep learning based adversarial attacks \cite{adversarial_attack_CL_2019, adversarial_attack_ICC_2018} are increasingly detrimental to communication security. This situation is more challenging in resource-constrained IoT applications since simple hardware architectures and software protocols cannot provide advanced countermeasures. All the mentioned challenges in resource-constrained IoT applications indicate the development of a new physical layer security framework.

This work proposes a waveform-defined security (WDS) framework, aiming to use a non-orthogonality concept in physical layer signal waveform to improve communication security. The fundamental waveform is based on \ac{SEFDM} \cite{SEFDM2003, TongyangTVT2017}, which intentionally creates \ac{ICI} via packing sub-carriers closer. The self-created interference complicates signal recovery but meanwhile increases the cost for eavesdropping. A similar concept was attempted by \cite{MOFDM_PIMRC2009} via overlapping two \ac{OFDM} signals to get a composite non-orthogonal signal. However, with the hardware advancement, brute-force \ac{ML} signal detection becomes realistic in low-cost hardware resulting in the security risk of eavesdropping.

Rather than a simple waveform design, the WDS framework designs a waveform tuning mechanism aiming to confuse eavesdroppers to misidentify signals. In this case, eavesdroppers can not recover signals even with the brute-force \ac{ML} detector. Therefore, only the legitimate user who knows exactly the signal format can recover signals. The work in \cite{Tongyang_CSNDSP_2020} initially revealed the feasibility of using the non-orthogonal SEFDM signal waveform in secure communications. However, further optimizations and practical experiments should be investigated to comprehensively verify the robustness of the framework.

The main contributions of this work are as the following.
\begin{itemize}
\item{ To deal with physical layer security in resource-constrained IoT, a WDS framework is proposed. WDS outperforms the recent PLS achievements in \cite{PLS_survey_2020}. Firstly, WDS avoids CSIT leading to a simpler solution compared to beamforming and artificial noise. Secondly, the non-orthogonal waveform structure in WDS adds \ac{ICI} leading to a more secure solution relative to \ac{OFDM}. Thirdly, WDS only modifies waveform and thus its hardware is simpler than MIMO and millimeter wave solutions. Fourthly, WDS supports omni-directional communications while \ac{NOMA} is limited to protected zone due to potential eavesdropping successive decoding. Finally, WDS has higher spectral efficiency than channel coding schemes via compressing occupied spectral bandwidth. } 

\item{ An efficient WDS waveform pattern is obtained after a joint study on three waveform tuning impact factors, namely training data feature, oversampling factor and \ac{BCF} offset.} 

\item{ The WDS framework can be easily incorporated into existing communication standards such as \ac{WLAN} IEEE 802.11a. A compatible WLAN-WDS frame is designed to preserve most of the original WLAN frame structure, which enables a straightforward deployment of the WDS framework in practice. }

\item{ A dual WDS security mechanism is designed. Firstly, the WLAN-WDS frame is so similar to the standard WLAN frame such that an eavesdropper would mistakenly decode WLAN-WDS frames using the standard WLAN protocol. Secondly, even the proper protocol is applied to decode WLAN-WDS frames, eavesdroppers cannot separate different signal patterns in the WDS framework, resulting in failed signal demodulation and detection. } 

\item{ Low-cost \ac{SDR} prototyping experiments are designed to validate the proposed WDS framework over the air. Experiments verify the feasibility of WDS framework in low-cost hardware and pave the way for applications in resource-constrained IoT scenarios. In addition, the experiments reveal the robustness of WDS framework even when eavesdroppers have advantages in signal power and channel conditions.} 

\end{itemize}

The rest of this paper is organized as follows. Section \ref{sec:waveform_fundamental} will introduce the fundamentals of the waveform. In Section \ref{sec:eavesdropper_model}, eavesdropping models applying maximum likelihood classifier, machine learning classifier and deep learning classifier, are investigated, followed by a brief description of signal detection. In Section \ref{sec:impact_factor_investigation}, three waveform tuning impact factors, namely training data feature, oversampling factor and \ac{BCF} offset, are studied to show their impacts on the WDS framework. A \ac{WLAN} coexistent scheme is studied in Section \ref{sec:WLAN_coexistence} showing the compatible integration of the WDS framework. A low-cost experiment is implemented in Section \ref{sec:SDR_experiment} to verify the feasibility of the proposed WDS framework in low-cost hardware, which further indicates its possibility in resource-constrained IoT applications. Finally, Section \ref{sec:conclusion} concludes the work.

\section{Waveform Fundamentals}\label{sec:waveform_fundamental}

\begin{figure}[t!]
\begin{center}
\includegraphics[scale=0.43]{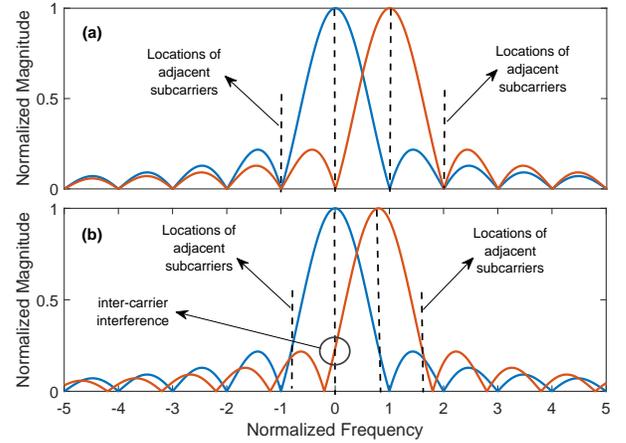}
\end{center}
\caption{Illustration of self-created inter carrier interference within SEFDM signal waveform. (a) OFDM sub-carrier packing. (b) SEFDM sub-carrier packing.}
\label{Fig:AI_encryption_subcarrier_packing}
\end{figure}

Non-orthogonal SEFDM waveform aims to compress signal spectral bandwidth while maintaining the same data rate. This is achieved by packing sub-carriers closer via breaking the orthogonality principle in \ac{OFDM}. The graphic explanation of SEFDM waveform is illustrated in Fig. \ref{Fig:AI_encryption_subcarrier_packing} where the same sub-carrier bandwidth is employed in OFDM and SEFDM except that the sub-carrier spacing in SEFDM is closer resulting in self-created \ac{ICI}.

A simplified mathematical format of one SEFDM symbol is expressed as
\begin{equation}
X_k=\frac{1}{\sqrt{Q}}\sum_{n=0}^{N-1}s_{n}\exp\left(\frac{j2{\pi}nk\alpha}{Q}\right),\label{eq:SEFDM_discrete_signal}\end{equation}
where the parameters are defined as

\begin{itemize}
\item{$X_k$, the time sample with the index of $k=0,1,...,Q-1$.  }
\item{$Q=\rho{N}$, the number of time samples.}  
\item{$N$, the number of sub-carriers.} 
\item{$\rho$, the oversampling factor.} 
\item{$\frac{1}{\sqrt{Q}}$, the scaling factor. } 
\item{$s_{n}$, the $n^{th}$ single-carrier symbol in one SEFDM symbol. } 
\item{$\alpha=\Delta{f}\cdotp{T}$, the bandwidth compression factor where $\Delta{f}$ is the sub-carrier spacing and $T$ is the time duration of one SEFDM symbol. } 
\end{itemize}

ICI will be introduced when $\alpha<1$. To mathematically show the impact of ICI, the instantaneous power for one SEFDM symbol, $X_k$, is computed as

\begin{equation}\label{eq:SEFDM_square_signal}
\begin{split}
|X_k|^2&=\frac{1}{Q}\sum_{n=0}^{N-1}\sum_{m=0}^{N-1}s_{n}s^{*}_{m}\exp\left(\frac{j2{\pi}(n-m)k\alpha}{Q}\right)\\
&=\underbrace{\frac{1}{Q}\sum_{n=0}^{N-1}|s_{n}|^2}_{Signal}+\\
&\underbrace{\frac{1}{Q}\sum_{n=0}^{N-1}\sum_{m\neq{n},m=0}^{N-1}s_{n}s^{*}_{m}\exp\left(\frac{j2{\pi}(n-m)k\alpha}{Q}\right)}_{ICI}.
\end{split}
\end{equation}

The signal power representation includes a signal term and an ICI term. When $\alpha=1$ for OFDM, the ICI term equals zero. However, for SEFDM signals with $\alpha<1$, the ICI term is not cancelled, which is the main factor that enables the non-orthogonal waveform a candidate for physical layer security. 

An \ac{IDFT} architecture is applicable to SEFDM signal generation. The general idea has been implemented in \ac{VLSI} \cite{PaulVLSI} and successfully applied in practical experiments \cite{TongyangTVT2017}. The basic principle is to pad zeros at the end of each vector $s$. Thus a longer symbol vector is achieved as
\begin{equation}
s^{'}_{n} = \left\{
  \begin{array}{l l}
    s_{n} & \quad \text{$0{\leq}n<N$}\\
    0 & \quad \text{$N{\leq}n<M$}
  \end{array}, \right.
\label{eq:symbol_vector_single_IFFT_SEFDM}\end{equation} 
where $M=Q/\alpha$ should be rounded to its closest integer. The direct modulation in \eqref{eq:SEFDM_discrete_signal} is therefore transformed to a typical \ac{IDFT} format as
\begin{equation}
X^{'}_k=\frac{1}{\sqrt{M}}\sum_{n=0}^{M-1}s^{'}_{n}\exp\left(\frac{j2{\pi}nk}M\right),\label{eq:signal_single_IFFT_pad_zeros}\end{equation}
where $n,k=[0,1,...,M-1]$. The output is truncated with only $Q$ samples reserved while the rest of the samples are discarded.

To simplify the expression, a matrix format of the signal generation is given by 
\begin{equation}
X=\mathbf{F}S=\underbrace{\mathbf{F^{'}}S^{'}}_{truncate},\label{eq:simple_SEFDM_signal}\end{equation}
where $X$ is a $Q$-dimensional vector of time samples, $S$ is an $N$-dimensional vector of transmitted symbols and $\mathbf{F}$ is a $Q\times{N}$ sub-carrier matrix with elements equal to $\exp({\frac{j2{\pi}{nk}\alpha}{Q}})$. $S^{'}$ is an $M$-dimensional vector of transmitted symbols and $\mathbf{F}^{'}$ is an $M\times{M}$ sub-carrier matrix with elements equal to $\exp({\frac{j2{\pi}{nk}}{M}})$. It is noted that the symbol vector obtained from the second multiplicative term should be truncated to $Q$ samples.

At the receiver side, the signal $Y$ is obtained via \ac{AWGN} channel as
\begin{equation}
Y=X+Z,\label{eq:simple_SEFDM_signal_AWGN}\end{equation}
where $Z$ is an $Q$-dimensional vector of noise samples. After signal demodulation via multiplying \eqref{eq:simple_SEFDM_signal_AWGN} with the conjugate sub-carrier matrix $\mathbf{F^{*}}$, an $N$-dimensional vector of demodulated symbols $R$ is obtained as
\begin{equation}
R=\mathbf{F^{*}}X+\mathbf{F^{*}}Z=\mathbf{F^{*}}\mathbf{F}S+\mathbf{F^{*}}Z=\mathbf{C}S+Z_{\mathbf{F^*}},\label{eq:correlation_FDM_signal}\end{equation}
where $\mathbf{C}$ is an $N\times N$ correlation matrix defined as $\mathbf{C}=\mathbf{F^{*}}\mathbf{F}$. To recover original signals $S$ from $R$, the \ac{ICI} caused by the correlation matrix $\mathbf{C}$ has to be mitigated using specially designed signal detectors.

Similar to the operations in \eqref{eq:symbol_vector_single_IFFT_SEFDM} and \eqref{eq:signal_single_IFFT_pad_zeros}, a \ac{DFT} architecture is applicable to SEFDM signal demodulation. After padding zeros at the end of $Y$, a longer symbol vector $Y^{'}$ is therefore obtained with its elements defined below

\begin{equation}
y^{'}_{n} = \left\{
  \begin{array}{l l}
    y_{n} & \quad \text{$0{\leq}n<Q$}\\
    0 & \quad \text{$Q{\leq}n<M$}
  \end{array}. \right.
\label{eq:symbol_vector_single_FFT_SEFDM}\end{equation} 
The M-point DFT for SEFDM signal demodulation is thus given by
\begin{equation}
R^{'}_k=\frac{1}{\sqrt{M}}\sum_{n=0}^{M-1}y^{'}_{n}\exp\left(\frac{-j2{\pi}nk}M\right),\label{eq:signal_single_FFT_pad_zeros}\end{equation}
where the output has to be truncated to $Q$ samples similar to the operation in \eqref{eq:signal_single_IFFT_pad_zeros}. In fact, to get the original symbol vector $S$, the output can be truncated to $N$ symbols directly.

\section{Eavesdropper Model} \label{sec:eavesdropper_model}

\begin{figure}[t!]
\begin{center}
\includegraphics[scale=0.37]{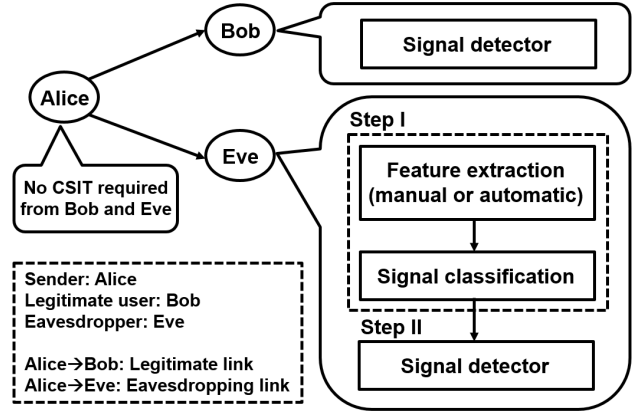}
\end{center}
\caption{Waveform defined secure communication.}
\label{Fig:WDS_eavesdropping_model}
\end{figure}

\begin{figure}[t!]
\begin{center}
\includegraphics[scale=0.37]{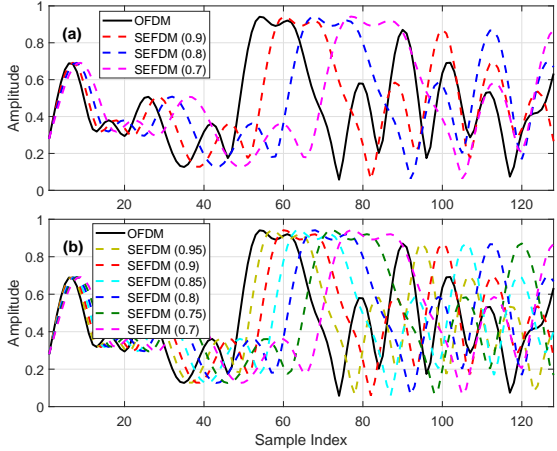}
\end{center}
\caption{Visualization of signal feature diversity and similarity via tuning SEFDM patterns. (a) Type-I signal pattern. (b) Type-II signal pattern. Values in the bracket indicate the bandwidth compression factor $\alpha$. To have a fair illustration comparison, the same QPSK single-carrier symbols are modulated by both Type-I and Type-II signals. In the rest of this work, random QPSK symbols will be used for each signal. }
\label{Fig:CNN_feature_combined_G_I_G_II}
\end{figure}

The proposed waveform based secure communication topology is presented in Fig. \ref{Fig:WDS_eavesdropping_model}. As commonly defined, Alice is the information sender, Bob is the legitimate user and Eve is the eavesdropper. The eavesdropper Eve is configured to be passive in this work, which only listens to signals and would not actively manipulate legitimate signals. As shown in Fig. \ref{Fig:WDS_eavesdropping_model}, there are two steps for a successful eavesdropping interception. Firstly, an eavesdropper should know the full information of signal formats. This could be achieved by manually or automatically extracting signal features followed by signal classification algorithms. Secondly, an efficient signal detector is required to recover signals based on the confirmed signal format from the first step. The specifically designed signal patterns, reused from \cite{tongyang_VTC2020_DL_classification}, are illustrated in Fig. \ref{Fig:CNN_feature_combined_G_I_G_II}. The Type-I signal pattern has strong signal diversity since the feature difference between adjacent signals is obvious. However, the Type-II signal pattern shows closer \ac{BCF} patterns and therefore strong signal feature similarity, resulting in more challenging signal classification at Eve. In terms of the legitimate communication link, the \ac{BCF} pattern is pre-known between Alice and Bob. Therefore, Bob will not need signal classification and will go through signal detection straightforwardly. {In practice, the \ac{BCF} pattern could be privately pre-shared between Alice and Bob. Another solution could design a BCF pattern generator that can reproduce identical BCF patterns at Alice and Bob. This work will consider pre-known BCF knowledge and skip the BCF pattern synchronization between Alice and Bob.}

{Each signal type has multiple BCF patterns. Therefore, in either one communication session or different communication sessions, BCF patterns will be dynamically changed. In this case, each transmitted symbol will use a different BCF configuration and this random-like BCF transmission strategy will confuse eavesdroppers and enhance the security level of WDS.}

It should be noted that Alice does not need any CSIT from Bob and Eve. The WDS secure communication framework is therefore less sensitive to channel environment variations and more robust than any other channel dependent physical layer security techniques. In addition, the avoidance of CSIT can simplify the entire system design, benefiting low-cost and resource-constrained communications.

\subsection{Learning and Classification Strategies}

Eavesdropping signal format classification models can be classified into maximum-likelihood based classifier, manual-feature based classifier and automatic-feature based classifier. {Unlike the commonly used \ac{RMSE} metric in regression models, the performance of a classification model will be measured by accuracy rate, which is the ratio of the number of correct classifications to the total number of classifications.}

The maximum-likelihood classifier provides an optimal solution. It was initially applied in modulation classification using single-carrier symbol-level likelihood functions \cite{classification_likelihood_AMC_TWC_2009, classification_likelihood_AMC_2011}. In an \ac{AWGN} channel with perfect knowledge of all parameters except the modulation format, the likelihood function is expressed as
\begin{equation}
L(r|\mathfrak{M},\sigma)=\frac{1}{P}\prod_{n=0}^{N-1}\sum_{p=0}^{P-1}\frac{1}{2\pi{\sigma^2}}\exp\left(-\frac{|r(n)-\mathfrak{M}(i,p)|^2}{2\sigma^2}\right),\label{eq:likelihood_function}\end{equation}
where $\mathfrak{M}$ indicates modulation candidates, $\mathfrak{M}(i,p)$ represents the $p^{th}$ constellation symbol in the $i^{th}$ modulation scheme. Each modulation scheme has up to $P$ constellation symbols. $N$ is the number of symbols for each observation, which indicates the number of sub-carriers in multicarrier signals. $\sigma^2$ is noise variance and $r(n)$ is the $n^{th}$ single-carrier complex symbol.

The maximum-likelihood classification is to maximize the likelihood function among all the modulation candidates. Assuming the entire potential solution set is $\Theta$, the maximum likelihood based solution $\hat{\mathfrak{M}}$ is give by
\begin{eqnarray}
\hat{\mathfrak{M}}=\arg\max_{\mathfrak{M}(i)\in \Theta}L(r|\mathfrak{M},\sigma).\label{eq:maximum-likelihood-decision}\end{eqnarray}

It is clearly seen from \eqref{eq:likelihood_function} and \eqref{eq:maximum-likelihood-decision} that the maximum-likelihood classification is limited to single-carrier symbols. It is also well noticing that this work focuses on signal format classification rather than modulation classification. The latter one can straightforwardly use the optimal maximum-likelihood function. However, signal format classification is more complex since most of the signals are based on multi-carrier structures. Without an accurate signal format classification as the first step, the subsequent modulation classification \cite{classification_likelihood_AMC_TWC_2009, classification_likelihood_AMC_2011} will not be achievable. To convert a multi-carrier signal into its baseband single-carrier symbols, the multi-carrier signal format has to be known, which is the challenge to be solved in this section.

The optimality of non-orthogonal signal classification has not been mathematically achieved since the conventional maximum-likelihood function is not applicable to multicarrier signals. In addition, the continuous variations of BCF values can theoretically lead to infinite classification solutions. Therefore, this section investigates two alternatives, which can classify multicarrier signals using manual-feature machine learning and automatic-feature deep learning.

Manual-feature classifiers rely on manual feature extractions followed by traditional machine learning classification methods. Professional domain-knowledge has to be applied to manually extract features, which could be time-domain characteristics, frequency-domain characteristics, wavelet transformed time-frequency characteristics and other statistical characteristics. Commonly used machine learning algorithms for classification tasks are \ac{SVM}, \ac{KNN}, decision trees, naive Bayes and neural networks. {Feature engineering is required to manually extract signal features at the first step. Previous work \cite{Tongyang_Globecom_wavelet_2020} revealed that wavelet time-frequency features with statistical feature dimensionality reduction schemes achieve the optimal classification accuracy. Therefore, the manual-feature classifier in this work will collect the dimensionality reduced wavelet time-frequency features for \ac{SVM} classification.}

\begin{figure}[t!]
\begin{center}
\includegraphics[scale=0.31]{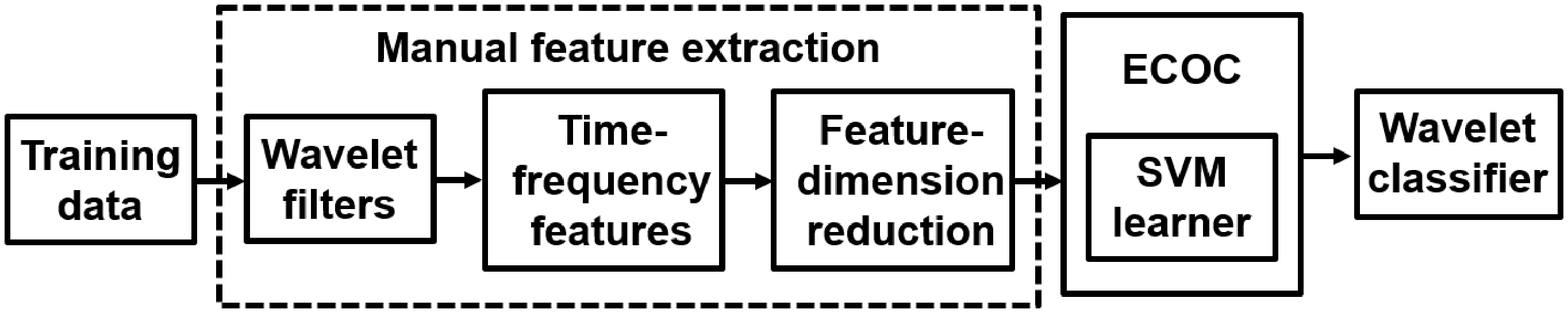}
\end{center}
\caption{Wavelet classifier training framework for the non-orthogonal signal classification.}
\label{Fig:WDS_wavelet_classifier_training_model}
\end{figure}

The wavelet classifier training framework in this work is presented in Fig. \ref{Fig:WDS_wavelet_classifier_training_model} where time-frequency features are obtained after using multiple wavelet filters on the training data. Since the extracted time-frequency features are two-dimensional, a statistical based dimension reduction scheme is applied to convert the two-dimensional feature matrix into a one-dimensional feature vector. As verified by \cite{Tongyang_Globecom_wavelet_2020}, the most efficient dimension reduction scheme relies on variance-interquartile-range statistical features. At the end, the \ac{ECOC} model with \ac{SVM} learners are used to train the wavelet classifier on the one-dimensional feature vector. It is clear that professional knowledge is required for the manual feature extraction process, which might be challenging for non-experts in this area. However, the manual feature extraction scheme will be more beneficial when the training data size is limited.

Automatic-feature classification, relying on deep learning, is becoming a popular approach to operate eavesdropping attack \cite{adversarial_attack_CL_2019, adversarial_attack_ICC_2018} since deep learning can simplify the training process without any professional domain knowledge for feature extractions. However, a large amount of data will be required by deep learning to automatically learn signal features and output a signal classifier model. A representative deep learning based classifier is \ac{CNN} \cite{OShea_classification_2018}, in which it proved 20\% higher modulation classification accuracy than traditional baseline classifiers. There are some other commonly used deep learning classification algorithms such as \ac{LSTM} and Autoencoder. However, they both have limitations in flexible feature extractions. CNN is a general deep learning method for classification tasks. It was initially applied in image classification and was later used in communication signals since a complex signal can be converted into a two-dimensional image with separate real and imaginary signal parts. A deep CNN would consist of several convolutional layers that will be used to automatically learn hidden features in a more flexible way than LSTM and Autoencoder. In addition, the implementation of CNN is more efficient since multiple nonlinear filters can work in parallel. Therefore, this work will focus on the CNN method and skip other deep learning algorithms. Previous work \cite{tongyang_VTC2020_DL_classification} designed an efficient CNN classifier for different non-orthogonal signals. In this work, the neural network architecture will be reused to test the robustness of the proposed WDS framework.

\begin{figure}[t!]
\begin{center}
\includegraphics[scale=0.238]{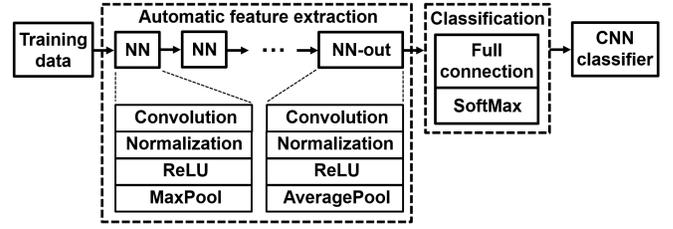}
\end{center}
\caption{CNN classifier training framework for the non-orthogonal signal classification.}
\label{Fig:WDS_CNN_classifier_training_model}
\end{figure}

The CNN neural network layer architecture in this work is presented in Fig. \ref{Fig:WDS_CNN_classifier_training_model}, in which multiple \ac{NN} modules are stacked to automatically learn signal features. There are additional sub-layers in each \ac{NN} module, namely convolutional layer, Batch-normalization layer, ReLU layer and MaxPool layer. The last NN module is responsible for outputting features. Therefore, an AveragePool layer is applied instead of the MaxPool layer. The classification is operated using a full-connection neural layer with SoftMax outputs. The optimal CNN classifier will be trained and updated iteratively via minimizing the cross-entropy loss between predicted values and true values.

It should be noted that the investigated IoT system in this work only includes legitimate users Alice and Bob. Since the eavesdropper Eve is not a part of the IoT system, its computational complexity is not considered in this work. In addition, the aim of Eve is to exhaustively identify legitimate user signals. In this case, both the machine learning and deep learning algorithms in this work are applied only at Eve. Therefore, the computational complexity of intelligent signal classifiers has nothing to do with the IoT system.

\subsection{Signal Detection}\label{subsec:signal_detection}

Signal detection is the second step of a complete eavesdropping, which aims to recover signals from ICI once signal formats are confirmed from the first step. The optimal detection algorithm is \ac{ML}, which searches all possible solutions and determines the optimal one as
\begin{eqnarray}{S}_{_{ML}}=\arg\min_{{{S}}\in O^N}\left\Vert {R}-\mathbf{C}{{S}}\right\Vert ^{2},\label{eq:WDS_ML}\end{eqnarray} 
where $O$ is the constellation cardinality and $O^N$ indicates the entire candidate solutions. By narrowing the search space, limited by a sphere radius $g$, the ML solution in \eqref{eq:WDS_ML} can be simplified into the \ac{SD} solution \cite{SD_orginal} as  
\begin{eqnarray}{S}_{_{SD}}=\arg\min_{{{S}}\in O^N}\left\Vert {R}-\mathbf{C}{{S}}\right\Vert ^{2}\label{eq:sd}\le{g},\end{eqnarray}
\begin{eqnarray}{g}=\left\Vert R-\mathbf{C}{S_{_{ZF}}}\right\Vert ^{2}\label{eq:SD_radius},\end{eqnarray}
where $S_{_{ZF}}=\lfloor{\mathbf{C^{-1}}R}\rceil$ is a coarse solution based on \ac{ZF}, which is used here to narrow the entire search space to a partial search space.

The \ac{SD} detector is simplified relative to \ac{ML} and it has been proved feasible in SEFDM signals with a large number of sub-carriers \cite{Tongyang_wincom2017} and strong ICI. However, its computational complexity is random \cite{KanarasCL2010} and is highly related to noise power. Therefore, the work in \cite{TongyangCL2013} proposed a simplified interference cancellation method named \ac{ID}, which can efficiently recover signals with minor ICI when the value of $\alpha$ is approaching one. The iterative cancellation is defined as
\begin{eqnarray}\label{eq:WDS_ID}
S_{\zeta}=R-(\mathbf{C}-\mathbf{e})S_{\zeta-1},
\end{eqnarray}
where $S_{\zeta}$ is the N-dimensional symbol vector after $\zeta$ iterations, $S_{\zeta-1}$ indicates the results after $\zeta-1$ iterations and $\mathbf{e}$ is an $N\times{N}$ identity matrix.

In summary, the \ac{ID} detector shows lower computational complexity but it is limited to signals with weak ICI \cite{TongyangCL2013}. When signals have strong ICI (i.e. small $\alpha$), the random-complexity \ac{SD} has to be used. Therefore, the choice of a signal detector depends on signal conditions, which is determined by the value of \ac{BCF} $\alpha$.

\section{Impact Factor Investigations} \label{sec:impact_factor_investigation}

This section evaluates three impact factors, which will determine the waveform characteristics and therefore affect eavesdropping signal classification accuracy. {Channel and hardware impairments are considered for training datasets in this work. A three-path wireless channel \ac{PDP} with path delay (s) [0 9e-6 1.7e-5] and path relative power (dB) [0 -2 -10] are reused from \cite{tongyang_VTC2020_DL_classification, Tongyang_Globecom_wavelet_2020, OShea_classification_2018, OShea_rml_datasets_2016}. The K-factor is 4 and the frequency offset is configured to be 2 \ac{PPM}. The maximum Doppler frequency is set to 4 Hz considering indoor people walking speed. In addition, the training dataset will cover a wide range of Es/N0 from -20 dB to 50 dB. Signal specifications are flexible and will be detailed in each scenario below.}

\subsection{Impact of Training Data} \label{subsec:training_data_effect}

In the previous work \cite{tongyang_VTC2020_DL_classification}, training dataset is generated based on a data augmentation (DA) principle. {This dataset generation mechanism aims at data-limited scenarios. The basic idea is to generate one data symbol, which will go through different feature-diversified wireless channels. The DA method will output multiple channel impaired symbols as a training dataset. Therefore, a symbol could be easily expanded to a large size dataset via time-variant wireless channels.} Although the DA based dataset generation has wireless channel diversity, it has limited signal feature diversity. Such a diversity-limited dataset is efficient in machine learning based classifier training \cite{Tongyang_Globecom_wavelet_2020} where features are manually extracted using expert knowledge. However, it might not be efficient for deep learning based \ac{CNN} classifier training \cite{tongyang_VTC2020_DL_classification} where a large amount of diversified data has to be used to automatically extract features. 

The DA based training methodology is evaluated at the beginning. {A source symbol is generated per signal class (i.e. per $\alpha$). Each OFDM or SEFDM symbol will have 2048 time samples via oversampling 256 single-carrier QPSK symbols by a factor of $\rho$=8 \cite{tongyang_VTC2020_DL_classification, Tongyang_Globecom_wavelet_2020}.} Each source symbol within a signal class will be expanded to 2,000 OFDM/SEFDM symbols {by going through independent time-variant wireless channels defined at the beginning of this section.} Therefore, the Type-I signal pattern will have four independent datasets consisting of overall 8,000 OFDM/SEFDM symbols. The same data generation principle is repeated for the Type-II signal pattern leading to seven datasets and overall 14,000 OFDM/SEFDM symbols. Based on the study in \cite{Tongyang_Globecom_wavelet_2020}, the value of Es/N0 has great effects on the training efficiency where a dataset covering a wide Es/N0 range would train a high accuracy classifier. Therefore, prior to the classifier training, the raw dataset would be contaminated by \ac{AWGN} ranging from Es/N0=-20 dB to 50 dB with an increment step of 10 dB. Such a dataset with rich \ac{AWGN} information would help to train a robust classifier.

\begin{figure}[t!]
\begin{center}
\includegraphics[scale=0.8]{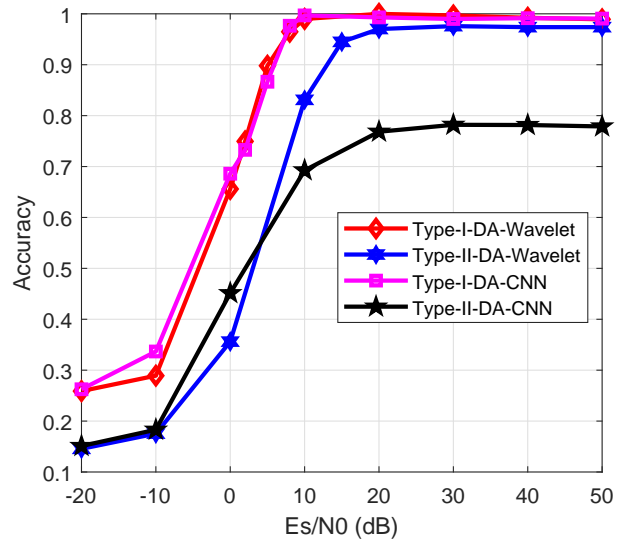}
\end{center}
\caption{Classification accuracy comparisons based on the data augmentation (DA) training methodology.}
\label{Fig:WDS_data_type_DA}
\end{figure}

To evaluate the classification accuracy at eavesdroppers, the wavelet classifiers are trained based on \cite{Tongyang_Globecom_wavelet_2020} in Fig. \ref{Fig:WDS_wavelet_classifier_training_model} and the CNN classifiers are trained according to \cite{tongyang_VTC2020_DL_classification} in Fig. \ref{Fig:WDS_CNN_classifier_training_model}. The classification accuracy, based on the DA training data, is presented in Fig. \ref{Fig:WDS_data_type_DA}. It should be noted that previous works in \cite{tongyang_VTC2020_DL_classification, Tongyang_Globecom_wavelet_2020} trained classifiers using the DA training data as well. Therefore, Fig. \ref{Fig:WDS_data_type_DA} summarizes what has been achieved in previous works and will be used as benchmarks for this work. It is clearly seen that both the CNN and wavelet Type-I signal classifiers perform well and reach 100\% accuracy. For the Type-II signal pattern, the wavelet classifier maintains similar accuracy but the accuracy of the CNN classifier drops by at least 20\%. This is expected because a diversity-limited dataset cannot efficiently assist CNN to automatically extract rich signal features while the manual feature extraction based wavelet classifier is not sensitive to the limited data diversity. It indicates that the DA based data generation is more suitable to wavelet classifier training.

To train a robust classifier for non-orthogonal waveforms, data diversity (DD) is enhanced via diversifying source data generation instead of the data augmentation. {The basic principle for DD is to generate multiple source symbols instead of a single source symbol in DA. Unlike the dataset expansion mechanism in DA, the enhanced feature of DD is that each source symbol will go through an independent channel. Therefore, the generated training dataset will have diversity both in source symbols and channel environments.} This subsection will generate 2,000 OFDM/SEFDM source symbols per signal class and each source symbol will be distorted by an independent time-variant wireless channel. Therefore, each signal class will have 2,000 random OFDM/SEFDM symbols with random wireless channel distortions. In this case, both data and channel characteristics have diversity and would be fair to both machine learning and deep learning classifier training.

\begin{figure}[t!]
\begin{center}
\includegraphics[scale=0.8]{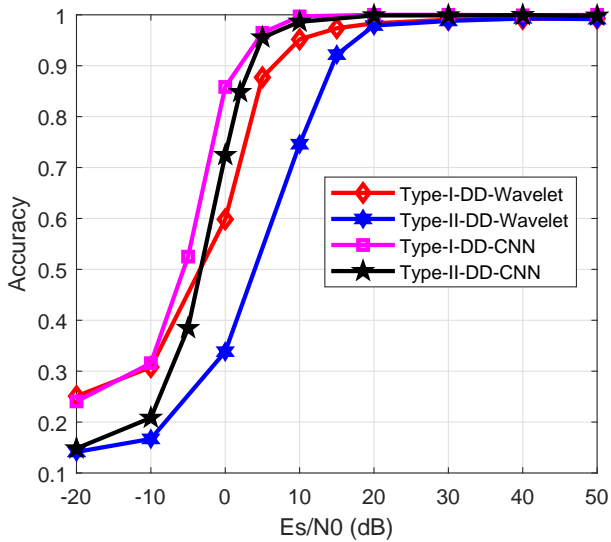}
\end{center}
\caption{Classification accuracy comparisons based on the data diversity (DD) training methodology.}
\label{Fig:WDS_data_type_DD}
\end{figure}

With the diversity enhanced dataset, both CNN classifiers and wavelet classifiers are re-trained following the same process in \cite{tongyang_VTC2020_DL_classification} and \cite{Tongyang_Globecom_wavelet_2020}, respectively. The results in Fig. \ref{Fig:WDS_data_type_DD} reveal that both the CNN and wavelet classifiers can reach 100\% classification accuracy at high Es/N0. It indicates that the DD based data generation is suitable to both CNN and wavelet classifiers training. Moreover, the CNN classifiers can even work well at low Es/N0. The significant achievement is at Es/N0=0 dB where the CNN based Type-II signal classification (72\% accuracy rate) has approximately 38\% higher accuracy than that of the wavelet Type-II signal classification (34\% accuracy rate).

In summary, the typical machine learning based wavelet classification method is robust when training data is generated via augmenting a limited dataset. The deep learning based CNN classifier is however sensitive to training data and its classification accuracy drops with a diversity-limited training dataset. With a diversity enhanced dataset, both the CNN and wavelet based classifiers can identify Type-I and Type-II signals at 100\% accuracy. Therefore, in the following, the DD based training data is used for both the CNN and wavelet classifiers while the DA based training data is merely for the wavelet classifiers.

\subsection{Impact of Oversampling} \label{subsec:oversampling_impact}

Oversampling factor $\rho$ in wired/wireless communications determines signal resolution. A higher value of $\rho$ leads to a better signal resolution. In addition, it is also a method to introduce spectral protection guard band \cite{Erik_book_4G}. Previous works \cite{tongyang_VTC2020_DL_classification, Tongyang_Globecom_wavelet_2020} followed a large oversampling factor of $\rho$=8, which is more than the requirements in practical systems such as 4G-LTE \cite{Erik_book_4G}, 5G-NR \cite{Erik_book_5G} and \ac{WLAN} 802.11 \cite{IEEE_802_11}. Therefore, this section will study the oversampling impact on non-orthogonal signal classification.

An oversampling factor determines the number of samples per symbol, which is expected to have more impacts on SEFDM signals as shown in \eqref{eq:SEFDM_square_signal}. It is clearly seen that the common signal term in either OFDM ($\alpha=1$) or SEFDM ($\alpha<1$) is only related to the raw single-carrier symbol $s_n$, which is independent from the oversampling factor $\rho$. The exponential term, $\exp({\frac{j2{\pi}{(n-m)k}\alpha}{Q}})$, in the ICI part, is zero for OFDM when $\alpha=1$. However, the term is not zero in SEFDM, which is determined by the factor $\rho$ because of $Q=\rho{N}$. Therefore, the value of $\rho$ will affect the accurate ICI expression and further determine the resolution of an SEFDM signal representation.

\begin{figure}[t!]
\begin{center}
\includegraphics[scale=1.05]{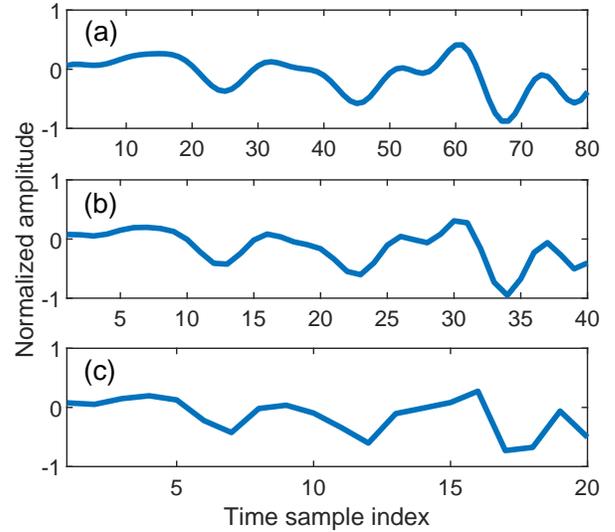}
\end{center}
\caption{Time-domain SEFDM sample ($\alpha$=0.8) illustration to show the impact of oversampling on the same data. (a)$\rho$=8. (b)$\rho$=4. (c)$\rho$=2.}
\label{Fig:WDS_time_domain_signal_oversampling_rho}
\end{figure}

To have a general idea of the oversampling impact on SEFDM, a set of integer values, $\rho=8, 4, 2$, are evaluated in Fig. \ref{Fig:WDS_time_domain_signal_oversampling_rho}. For the purpose of illustration, a total number of 20 time samples are truncated for the case of $\rho$=2. Based on this benchmark, a number of 40 time samples are truncated for the case of $\rho$=4 and 80 time samples for $\rho$=8.

The SEFDM signal of $\alpha$=0.8, with an oversampling factor of $\rho$=8, is illustrated in Fig. \ref{Fig:WDS_time_domain_signal_oversampling_rho}(a). The time-domain sample waveform is smooth and indicates a sufficient signal resolution. With the reduction of an oversampling factor to $\rho$=4, the SEFDM signal resolution is reduced and its time-domain waveform is slightly distorted in Fig. \ref{Fig:WDS_time_domain_signal_oversampling_rho}(b). Further reducing the oversampling factor results in Fig. \ref{Fig:WDS_time_domain_signal_oversampling_rho}(c). It is clearly seen that the signal shape is greatly changed and the signal profile is not following the one in Fig. \ref{Fig:WDS_time_domain_signal_oversampling_rho}(a). It is inferred that the reduction of an oversampling factor would have apparent effects on accurate SEFDM signal representations.

\begin{figure}[t!]
\begin{center}
\includegraphics[scale=0.8]{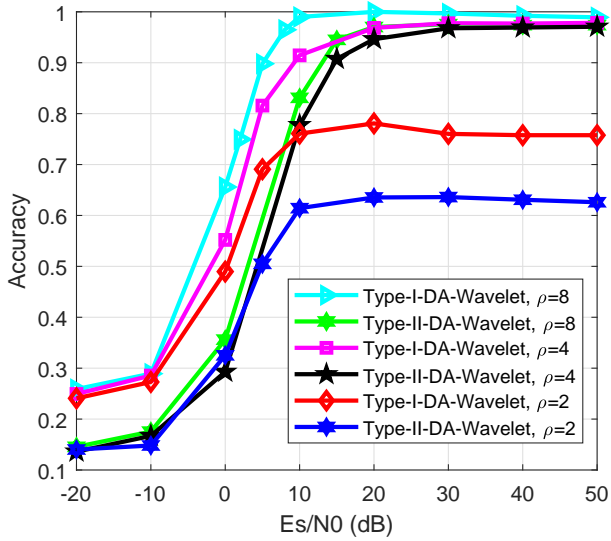}
\end{center}
\caption{Wavelet classification accuracy comparisons with various oversampling factors on the DA based training dataset.}
\label{Fig:WDS_rho_DA_8_4_2_wavelet}
\end{figure}

For Type-I and Type-II signals of $\rho$=8, 4, the DA based wavelet classification accuracy is between 95\% and 100\% as presented in Fig. \ref{Fig:WDS_rho_DA_8_4_2_wavelet}. With a further reduced oversampling factor to $\rho$=2, the Type-I signal classification accuracy drops to 75\% while the Type-II accuracy drops to 63\%. This is due to the reduced SEFDM signal resolutions and therefore inaccurate ICI representations.

\begin{figure}[t!]
\begin{center}
\includegraphics[scale=0.8]{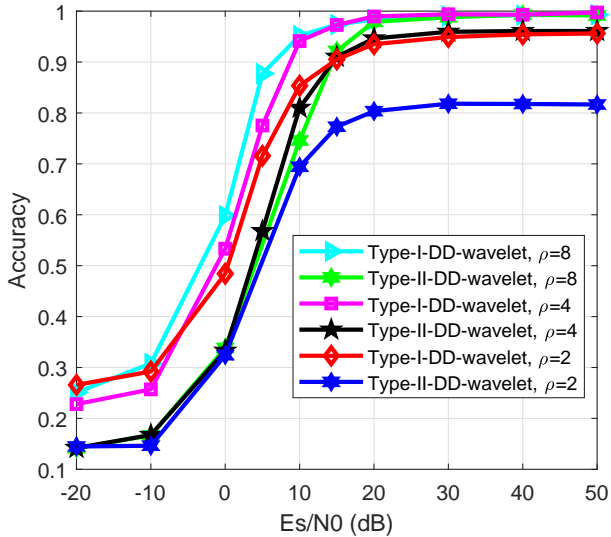}
\end{center}
\caption{Wavelet classification accuracy comparisons with various oversampling factors on the DD based training dataset.}
\label{Fig:WDS_rho_DD_8_4_2_wavelet}
\end{figure}

The DD based wavelet classifier is tested with results showing in Fig. \ref{Fig:WDS_rho_DD_8_4_2_wavelet}. A similar trend is observed for the cases of $\rho$=8, 4, where the accuracy is between 95\% and 100\%. However, data diversity can efficiently improve the classification accuracy for the case of using a small oversampling factor $\rho$=2. Results show that the Type-I signals can be identified with 95\% accuracy and the Type-II signals with 82\% accuracy. Comparing with the same oversampled signals in Fig. \ref{Fig:WDS_rho_DA_8_4_2_wavelet}, the accuracy rates for the Type-I and Type-II signals are improved approximately by 27\% and 30\%, respectively. The results indicate that both the DD and DA training methodologies achieve similar classification performance when the oversampling factor $\rho$ is large sufficient. However, when the oversampling factor $\rho$ is small, the DD based training is more robust than the DA based training. Therefore, in the following studies, the DD training method will be used for wavelet classifiers training.

\begin{figure}[t!]
\begin{center}
\includegraphics[scale=0.8]{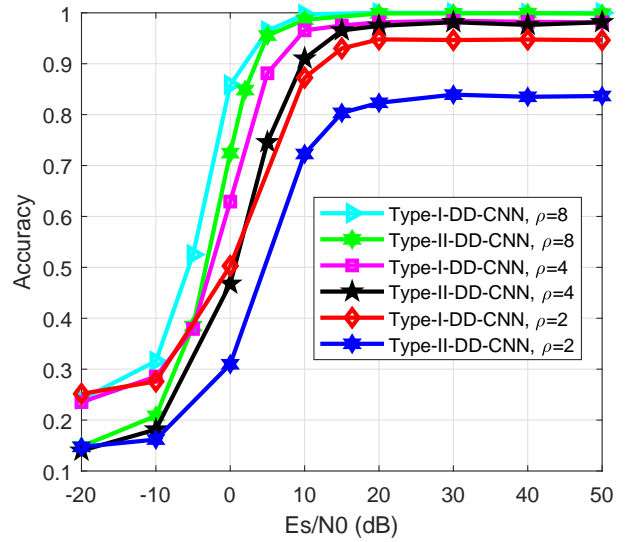}
\end{center}
\caption{CNN classification accuracy comparisons with various oversampling factors on the DD based training dataset.}
\label{Fig:WDS_rho_DD_8_4_2_CNN}
\end{figure}

The CNN classifiers, trained by signals with different values of $\rho$, are evaluated in Fig. \ref{Fig:WDS_rho_DD_8_4_2_CNN}. The signals with large oversampling factors, $\rho$=8, 4, can achieve high accuracy between 97\% and 100\%. When the factor is reduced to $\rho$=2, the Type-I signal maintains a high accuracy at 95\% while the Type-II signal accuracy drops to around 84\%.

In summary, oversampling determines the number of samples used to represent one symbol and therefore determines the resolution of a signal. The level of oversampling affects accurate feature extractions and robust classifier training especially for non-orthogonal SEFDM signals. A large oversampling factor leads to a better training condition and therefore higher classification accuracy but at the cost of time sample redundancy. A small oversampling factor will save time sample resources, which is more realistic in practical communication systems. It should be noted that the accuracy degradation due to reduced oversampling will be an extra benefit to SEFDM signals in terms of secure communications. For a comprehensive comparison, an upper-bound case of $\rho$=8 (following \cite{tongyang_VTC2020_DL_classification,Tongyang_Globecom_wavelet_2020}) and a lower-bound case of $\rho$=2 will be both considered in the following studies.

\subsection{Impact of BCF Offset} \label{subsec:BCF_gap}

Tuning the bandwidth compression factor will modify signal feature similarity and diversity. Previous work studied two types of signal patterns in \cite{tongyang_VTC2020_DL_classification}. As shown in Fig. \ref{Fig:CNN_feature_combined_G_I_G_II}, the Type-I signals with BCF offset $\Delta\alpha$=0.1 have a signal pattern of $\alpha$=1.0, 0.9, 0.8, 0.7. The Type-I signals are easily identified by either CNN classifiers or wavelet classifiers. A more challenging scenario is the Type-II signal pattern where its \ac{BCF} offset is $\Delta\alpha$=0.05, which results in a feature-similarity dominant scenario. With proper dataset training and symbol oversampling, the Type-II signals can be classified by both CNN classifiers and wavelet classifiers with results shown in Section \ref{subsec:training_data_effect} and Section \ref{subsec:oversampling_impact}.

\begin{figure}[t!]
\begin{center}
\includegraphics[scale=0.8]{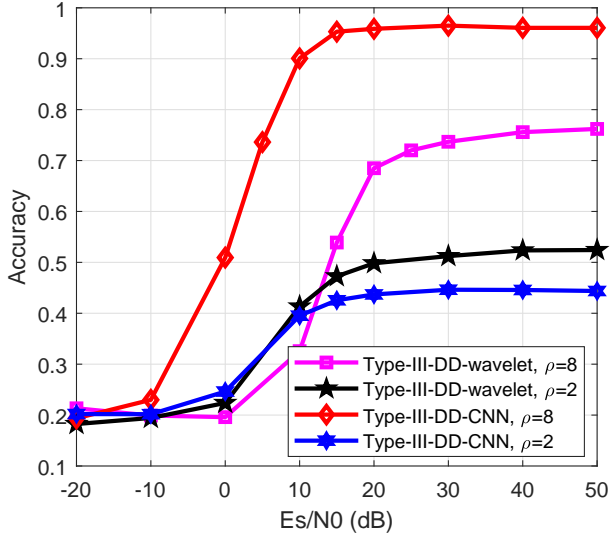}
\end{center}
\caption{Classification accuracy for the Type-III signal pattern.}
\label{Fig:WDS_Type_III_rho_DD_8_2_CNN_wavelet}
\end{figure}

An inspiring enhancement approach for communication security is to narrow the \ac{BCF} offset further and to have a more challenging feature-similarity dominant scenario. The tuning of BCF offset is flexible and has a number of patterns. This section will choose one pattern for an example demonstration. The optimal tuning pattern is not extensively investigated in this work. A signal pattern with BCF offset $\Delta\alpha$=0.015, termed Type-III, is designed in this section with bandwidth compression factors $\alpha$=1, 0.985, 0.97, 0.955, 0.94. Fig. \ref{Fig:WDS_Type_III_rho_DD_8_2_CNN_wavelet} will evaluate the Type-III signal pattern classification accuracy using the DD based training method and with $\rho$=8, 2.

It is clearly seen in Fig. \ref{Fig:WDS_Type_III_rho_DD_8_2_CNN_wavelet} that with the sufficient oversampling $\rho$=8, the CNN classifier still achieves a high accuracy rate at around 96\%. With the same oversampling, the wavelet classifier can only reach 76\% accuracy. Reducing the factor to $\rho$=2, the accuracy rates of both the CNN and wavelet classifiers will drop to 52\% and 44\%, respectively.

In summary, a large value of BCF offset, such that in the Type-I and Type-II signal patterns, would simplify signal classification. However, a small value of BCF offset would challenge accurate signal identification. It is inferred that with a further reduction of BCF offset, an accurate signal classification would be impossible. In addition, oversampling has a greater effect on the Type-III signal pattern than the other two signal types.

\section{WLAN Coexistence} \label{sec:WLAN_coexistence}

\ac{WLAN} is a ubiquitous technique being used in our daily life. Emerging data-hungry IoT applications are increasingly dependent on WLAN networks, such as remote monitoring, in which a large amount of video/voice data is generated and might be uploaded to the cloud. Typical narrowband IoT techniques such as ZigBee, LoRa, SigFox and NB-IoT would not be possible to achieve this. In addition, most narrowband IoT applications would require WLAN gateways to connect to the Internet. Therefore, the importance of WLAN in IoT applications is significant.   

To show the coexistence capability of the WDS framework with existing communication systems, the Type-III signal pattern is integrated in the \ac{WLAN} standard following IEEE 802.11a signal specifications \cite{IEEE_802_11}. By simply upgrading the IFFT signal generation methodology, the proposed WDS framework can be deployed straightforwardly in existing WLAN systems. {It should be noted that this work considers coexistence of WDS frames and existing WLAN frames in a time-division multiplexing mode. Therefore, inter-band coexistence interference is not introduced in this work. For complex scenarios when WDS frames and WLAN frames are working in a frequency-division multiplexing mode, inter-band coexistence interference will appear. Previous work in \cite{Xinyue_6G_Coexistence} has studied its impact and research results verified that the inter-band interference can be mitigated by using coding schemes. Due to limited space, this work will merely consider the basic scenario when WDS frames and WLAN frames are working in a time-division multiplexing mode, which has no coexistence interference.}

\subsection{WDS Framework in IEEE 802.11a}

IEEE 802.11a has a unique signal structure, which consists of 48 data symbols and 4 pilot symbols. With a fractional oversampling factor of $\rho$=16/13, the IFFT size is 64. It is clear that the IFFT size in 802.11a is smaller than that of the previous simulations in Section \ref{sec:impact_factor_investigation}. The smaller oversampling factor will additionally enhance communication security as proved by Section \ref{subsec:oversampling_impact}.

As clarified in \eqref{eq:signal_single_IFFT_pad_zeros}, the typical SEFDM signal is implementable by IFFT. Therefore, the SEFDM signal generation is highly coexistent with the 802.11a standard via merely modifying the length of IFFT. At the receiver, an FFT is also applicable to match the transmitter side architecture. All other signal processing and system architectures maintain the same with 802.11a. In this case, SEFDM signal generation and reception are highly compatible with the standard \ac{WLAN} framework.

\begin{figure}[t!]
\begin{center}
\includegraphics[scale=0.35]{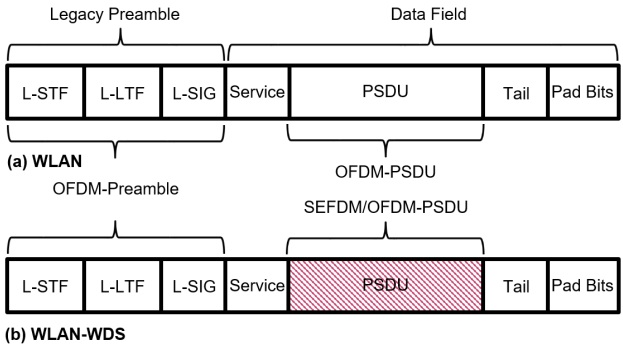}
\end{center}
\caption{The design of WDS signal frame coexisting with the standard WLAN 802.11a. (a) Typical WLAN 802.11a frame. (b) Proposed WLAN-WDS frame.}
\label{Fig:WDS_WLAN_frame}
\end{figure}

The 802.11a frame structure is presented in Fig. \ref{Fig:WDS_WLAN_frame}(a). The entire frame is termed physical layer conformance procedure (PLCP) protocol data unit (PPDU) \cite{IEEE_802_11}, which includes legacy preamble and data field. The legacy preamble, consisting of legacy short training field (L-STF), legacy long training field (L-LTF) and legacy signal (L-SIG) field, is used for frequency compensation, phase correction, timing synchronization, channel estimation, \ac{AGC} adjustment, \ac{MCS} notification, etc. The PLCP service data unit (PSDU) in the data field is responsible for carrying data symbols.

To have a high compatibility with 802.11a, only the PSDU field will be replaced by the Type-III pattern signals and any other fields will be maintained as the 802.11a standard. In this case, the modification to the existing 802.11a standard is minor. Unlike the conventional WLAN frame including OFDM-preamble and OFDM-PSDU, the newly designed WLAN-WDS frame will consist of OFDM-preamble and Type-III pattern based SEFDM/OFDM-PSDU.

Once a WLAN-WDS frame is received, an eavesdropper will have two possible actions to recover signals. In Scenario-I, Eve will mistakenly assume that all captured frames are defined by the traditional WLAN 802.11a standard as shown in Fig. \ref{Fig:WDS_WLAN_frame}(a). This makes sense because a WLAN-WDS frame reuses the WLAN legacy preamble as shown in Fig. \ref{Fig:WDS_WLAN_frame}(b), which would give a wrong indication that the following PSDU is also specified by 802.11a. Therefore, the mismatch between the information extracted from the legacy preamble and the PSDU data will confuse Eve to mistakenly use the typical 802.11a standard to recover the WLAN-WDS PSDU field even the data field PSDU is actually modulated by mixed SEFDM/OFDM symbols. The incorrect PSDU demodulation will significantly affect signal detection. In Scenario-II, the eavesdropper is assumed to realize the unique PSDU pattern in WLAN-WDS frames. However, the eavesdropper cannot easily recognize the difference between SEFDM-PSDU and OFDM-PSDU since an efficient signal classifier is not available. Therefore, the subsequent signal detection will not be reliable. 

A list of terms are defined below and will be used in the following simulation and experiment.

\begin{itemize}
\item{WLAN-OFDM: OFDM symbols are generated for PSDU following the 802.11a standard, which has 48 data symbols, 4 pilot symbols, 64 time samples and $\rho$=16/13.  } 
\item{WLAN-SEFDM: SEFDM symbols are generated for PSDU following the 802.11a standard, which has 48 data symbols, 4 pilot symbols, 64 time samples and $\rho$=16/13.  } 
\item{WLAN-WDS: the WDS framework is integrated in the standard WLAN frame, as shown in Fig. \ref{Fig:WDS_WLAN_frame}(b).} 
\item{WLAN-Type-III: the Type-III signal pattern, consisting of WLAN-OFDM and WLAN-SEFDM, is applied to the PSDU field in an WLAN-WDS frame.} 
\end{itemize}

\subsection{Reliability and Security} \label{subsec:reliability_security}

According to the available physical layer security experiment research in \cite{PLS_practical_CNS_mmWave_2015, PLS_practical_DM_2010, PLS_practical_GC_AN_2017, PLS_practical_VTC_coding_2018}, the commonly used security metrics are signal power difference and \ac{BER}. It is well noticing that the security enhancement of the proposed WDS framework is not dependent on the signal power advantage of legitimate communication links over eavesdropper communication links. The aim of the framework is to confuse eavesdroppers. Therefore, an eavesdropper will not break the communication security even its received signal power is higher than the legitimate user. In this case, instead of considering the signal power difference between legitimate users and eavesdroppers, this work will use BER as one metric to evaluate the communication security. Additionally, a specific metric, termed confusion matrix, is also applied in this work. Unlike the average accuracy results on entire signal classes in Section \ref{sec:impact_factor_investigation}, a confusion matrix can tell the details of classification for each signal class.

\begin{figure}[t!]
\begin{center}
\includegraphics[scale=0.8]{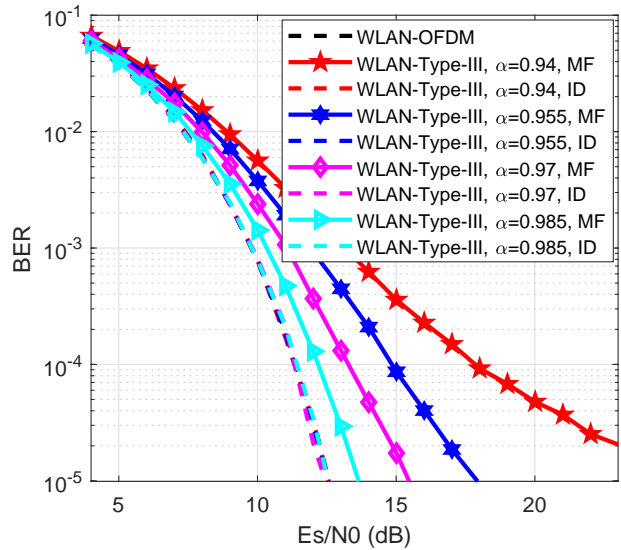}
\end{center}
\caption{BER performance of detecting WLAN-Type-III signals at the legitimate user.}
\label{Fig:WDS_WLAN_Type_III_BER_MF_ID}
\end{figure}

This section focuses on the feature-similarity dominant Type-III signal pattern. As explained in Fig. \ref{Fig:WDS_eavesdropping_model} that the legitimate user Bob knows accurate signal formats based on pre-shared information. Therefore, Bob can decode signals without the first-step signal classification. Based on the descriptions in Section \ref{subsec:signal_detection}, the simple ID signal detector is sufficient to recover non-orthogonal signals when the self-created \ac{ICI} is not strong. Since the minimum bandwidth compression factor in the Type-III signal pattern is $\alpha$=0.94, therefore the ID detector is sufficient. The BER performance for the signals at legitimate user Bob is presented in Fig. \ref{Fig:WDS_WLAN_Type_III_BER_MF_ID}. It is clearly seen that using the typical \ac{MF} detector, performance loss will exist and the signal with $\alpha$=0.94 has the worst performance. On the other hand, all the signals can be recovered perfectly by ID detectors leading to identical performance with the WLAN-OFDM signal. This indicates the WDS framework reliability at a legitimate user when the ID detector is applied.

In terms of BER performance at Eve, there are two possible eavesdropping scenarios considering the use of the proposed WLAN-WDS frame in Fig. \ref{Fig:WDS_WLAN_frame}(b). To test the BER performance, a total of 5,000 OFDM/SEFDM symbols are generated and each signal class has 1,000 symbols.

\begin{figure}[t!]
\begin{center}
\includegraphics[scale=0.8]{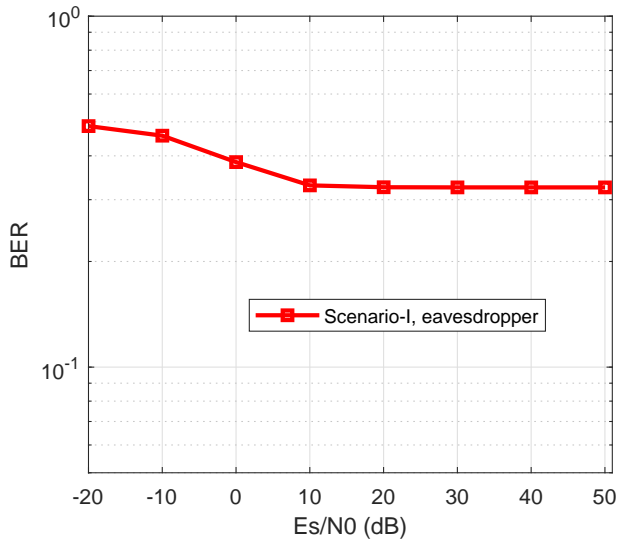}
\end{center}
\caption{Scenario-I: BER performance of detecting WLAN-Type-III signals at the eavesdropper when received symbols are incorrectly demodulated and detected following the WLAN-OFDM specification.}
\label{Fig:WDS_WLAN_Type_III_Eve_BER_CNN_wavelet_scenario_I}
\end{figure}

In Scenario-I, all the symbols will be incorrectly demodulated based on WLAN-OFDM specifications. The improper operation from Eve results in extremely degraded BER in Fig. \ref{Fig:WDS_WLAN_Type_III_Eve_BER_CNN_wavelet_scenario_I}. It is seen that the result of Scenario-I converges to a flat BER curve when Es/N0 is increased, which indicates a failure of eavesdropping.

\begin{table}[t!]
\caption{CNN classifier neural network layer architecture}
\centering
\begin{tabular}{ll}
\hline \hline
$\mathbf{Layers}$ & $\mathbf{Dimension}$  \\[0.5pt] \hline 
Input layer & $2\times{64}$ \\ 
Convolutional layer-1 & $2\times{64}\times{64}$ \\ 
Convolutional layer-2 & $2\times{32}\times{64}$ \\ 
Convolutional layer-3 & $2\times{16}\times{64}$ \\ 
Convolutional layer-4 & $2\times{4}\times{64}$ \\ 
Convolutional layer-5 & $2\times{2}\times{64}$ \\ 
Full-connection layer & $2\times{1}\times{64}$ \\ 
SoftMax output layer & $1\times{1}\times{5}$ \\\hline \hline
\label{tab:table_CNN_architecture_WLAN}
\end{tabular}
\end{table}

In Scenario-II, it is assumed that Eve will notice the tricks of the WLAN-WDS PSDU and will apply signal classifiers to identify each signal format. Following the same training methodology in Fig. \ref{Fig:WDS_CNN_classifier_training_model}, a CNN classifier is re-trained for the WLAN-Type-III signals with the neural network architecture presented in Table \ref{tab:table_CNN_architecture_WLAN} where five convolutional layers are stacked. A wavelet classifier is re-trained following Fig. \ref{Fig:WDS_wavelet_classifier_training_model} based on the manually extracted variance-interquartile-range features and the \ac{ECOC} model with \ac{SVM} learners.

\begin{figure}[t!]
\begin{center}
\includegraphics[scale=0.8]{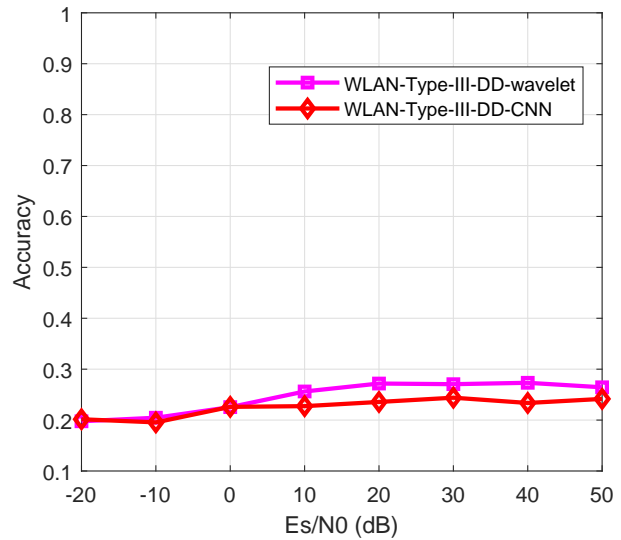}
\end{center}
\caption{WLAN-Type-III signal classification at the eavesdropper.}
\label{Fig:WDS_WLAN_Type_III_DD_CNN_wavelet}
\end{figure}

The classification accuracy for the WLAN-Type-III signal pattern is shown in Fig. \ref{Fig:WDS_WLAN_Type_III_DD_CNN_wavelet}. It is clearly seen that when applying WLAN-Type-III structured signals with an oversampling factor $\rho$=16/13, both the CNN and wavelet classifiers cannot identify signals properly with only 24\% and 27\% accuracy, respectively. The reduced accuracy is expected since Fig. \ref{Fig:WDS_Type_III_rho_DD_8_2_CNN_wavelet} has verified that a small oversampling factor will degrade classification accuracy.

\begin{figure}[t!]
\begin{center}
\includegraphics[scale=0.8]{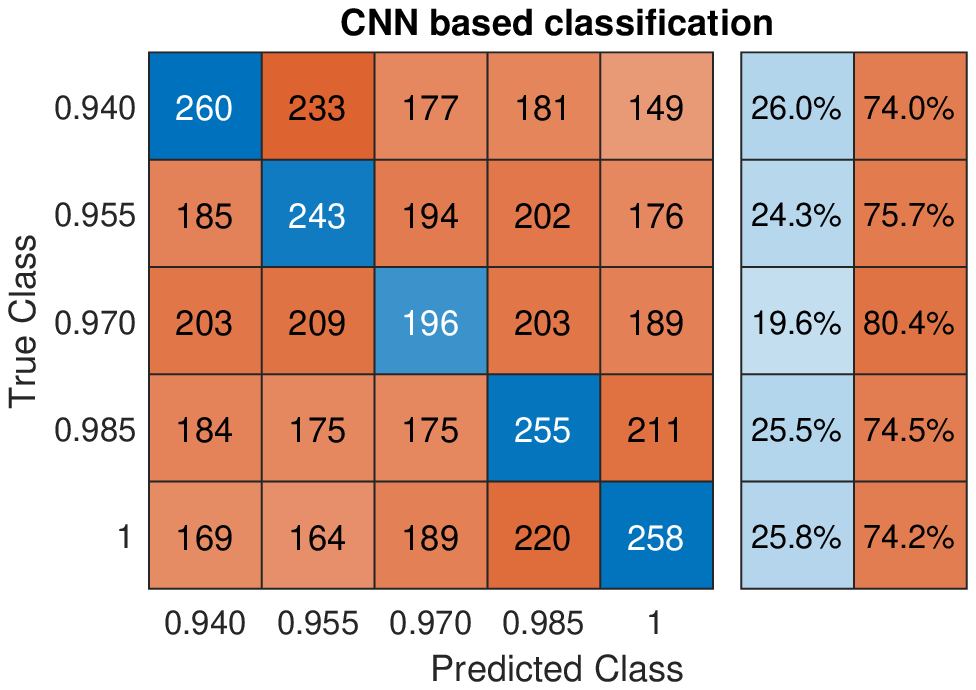}
\end{center}
\caption{Confusion matrix of WLAN-Type-III signals CNN classification at Es/N0=20 dB. $\alpha$=1 indicates WLAN-OFDM while other values of $\alpha$ indicate WLAN-SEFDM.}
\label{Fig:WDS_confusion_matrix_CNN}
\end{figure}

\begin{figure}[t!]
\begin{center}
\includegraphics[scale=0.8]{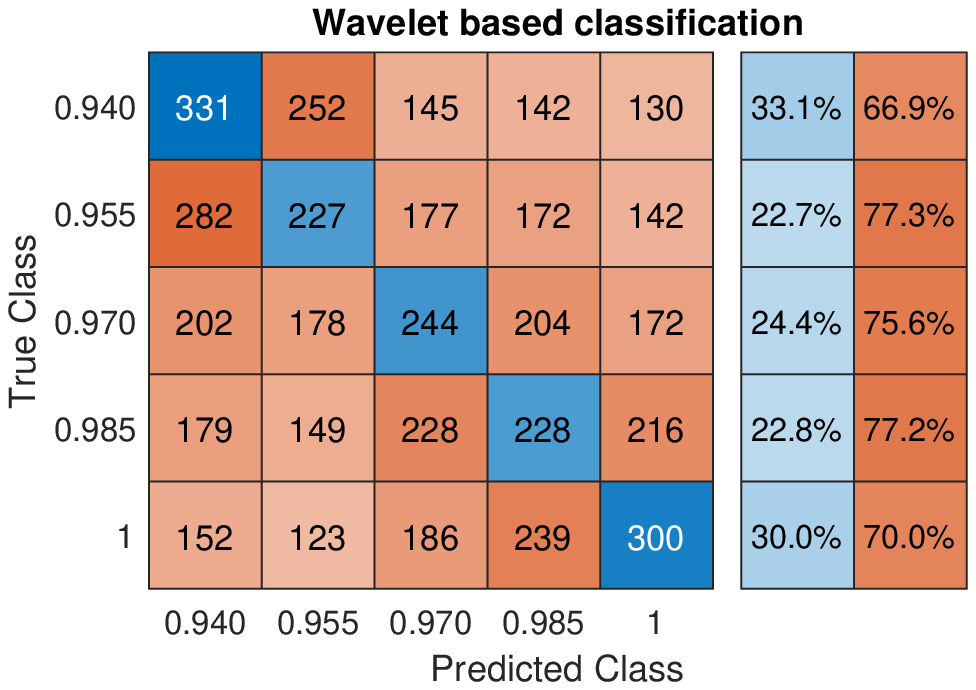}
\end{center}
\caption{Confusion matrix of WLAN-Type-III signals wavelet classification at Es/N0=20 dB. $\alpha$=1 indicates WLAN-OFDM while other values of $\alpha$ indicate WLAN-SEFDM.}
\label{Fig:WDS_confusion_matrix_Wavelet}
\end{figure}

The detailed results of classification can be expressed in the format of confusion matrix, in which diagonal elements indicate perfect classification while any non-diagonal elements indicate misclassification. {A confusion matrix commonly uses different coloured grids to give, at first glance, a general idea of correct classification and incorrect classification. However, a more scientific approach is to examine the values in each coloured grid. The values in each diagonal grid indicate the number of correctly identified symbols while other values indicate incorrectly identified symbols. In addition, a separate percentage table is jointly presented, in which the first column represents correct classification accuracy rates and the second column represents incorrect classification accuracy rates.} The CNN based confusion matrix is presented in Fig. \ref{Fig:WDS_confusion_matrix_CNN}. The values on the vertical axis indicate BCF for true signal classes while the horizontal axis shows predicted signal classes. The average accuracy for each signal class is between 19.6\% and 26\%. It is clearly seen that only 25.8\% of transmitted WLAN-OFDM signals are properly classified. A similar result for wavelet classification is illustrated in Fig. \ref{Fig:WDS_confusion_matrix_Wavelet} where the average accuracy is ranged from 22.7\% to 33.1\%. In terms of the WLAN-OFDM signal, its correct classification is at 30\%. Most of the misclassified WLAN-OFDM signals are concentrated in its adjacent signal class, $\alpha$=0.985. This is due to the fact that the WLAN-SEFDM signal of $\alpha$=0.985 is more similar to the WLAN-OFDM signal than any other WLAN-SEFDM signals. In this case, misclassification is unavoidable and communication security is beneficially enhanced.

\begin{figure}[t!]
\begin{center}
\includegraphics[scale=0.8]{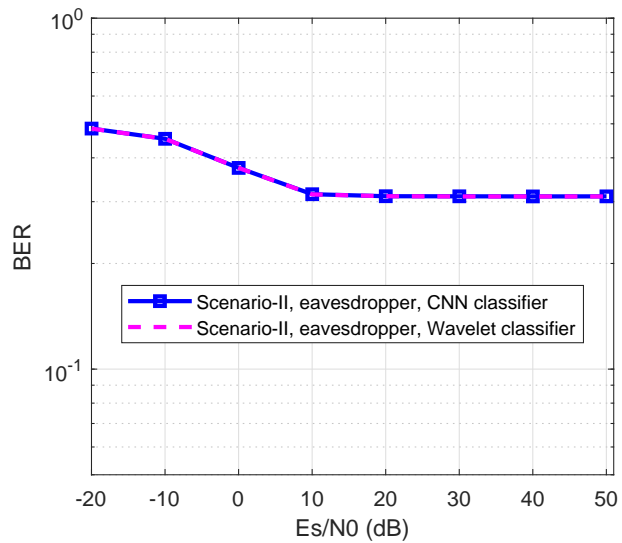}
\end{center}
\caption{Scenario-II: BER performance of WLAN-Type-III signals at the eavesdropper when received symbols are incorrectly demodulated and detected following the confusion matrix classification mapping scheme.}
\label{Fig:WDS_WLAN_Type_III_Eve_BER_CNN_wavelet_scenario_II}
\end{figure}

The performance evaluation of Scenario-II is more complex than that of Scenario-I. As shown in Fig. \ref{Fig:WDS_confusion_matrix_CNN} and Fig. \ref{Fig:WDS_confusion_matrix_Wavelet}, each confusion matrix has 25 possible classification mapping schemes. Each mapping, either correct or incorrect, will indicate one signal recovery scheme. Therefore, there are overall 25 signal recovery schemes per confusion matrix. One signal recovery scheme includes signal demodulation and signal detection. The function of one complete recovery process is to demodulate and detect true class labelled signals using predicted class labels. To get a general idea of the performance, a single BER result is obtained by averaging the 25 possible results per confusion matrix. Therefore, for either the CNN classifier or the wavelet classifier, a total of eight BER results will be obtained when considering Es/N0 from -20 dB to 50 dB with a 10 dB increment step.

Prior to the BER calculation, multiple confusion matrices should be obtained at Es/N0=-20 dB:50 dB. Due to the limited space, this work merely presents the confusion matrices at Es/N0=20 dB in Fig. \ref{Fig:WDS_confusion_matrix_CNN} and Fig. \ref{Fig:WDS_confusion_matrix_Wavelet}. The final BER of Scenario-II will consider all the confusion matrices Es/N0=-20 dB:50 dB leading to Fig. \ref{Fig:WDS_WLAN_Type_III_Eve_BER_CNN_wavelet_scenario_II}. The results show clearly that without an accurate classifier, which is currently unachievable, the eavesdropper cannot properly recover signals.

It reveals that the WDS framework in WLAN maintains the legitimate user side reliability using a specially designed signal detector. In addition, the joint study of confusion matrix and BER verifies that the framework enhances security by confusing an eavesdropper in two possible scenarios.

\subsection{Complexity Analysis}

The complexity of WDS framework should be evaluated comprehensively. It can be divided into signal processing complexity and hardware architecture complexity. 

WDS signal processing includes signal generation, signal classification and signal detection. Taking Fig. \ref{Fig:WDS_eavesdropping_model} as an example. Alice is responsible for signal generation. Bob will use the \ac{ID} detector to recover signals. Eve will need signal classification and signal detection. Since the purpose of Eve is to exhaustively recover signals without considering power/resource consumption, therefore the complexity at Eve is ignored in this work. {In principle, the higher eavesdropping computational complexity the better security level for the proposed WDS framework.}

Unlike typical WLAN communications where more traffic happens at downlink, IoT based applications would generate more traffic at uplink. In this case, the signal generation at Alice who is functioned as an IoT user, will be more concerned in this work. The signal detection at Bob, which would be at a WLAN router, is not considered. The reason is that WLAN routers are powered via wires and power consumption or computational complexity is not very crucial to the implementation of WDS. Therefore, the computational complexity in this section focuses merely at an IoT user side. In terms of the downlink channel from a WLAN router to an IoT device, the original 802.11a standard will be used in order to maintain simple signal processing at an IoT device. Since an IoT downlink channel is responsible for crucial control instructions that will determine the working principle for each IoT device, the WDS framework is also applicable but at the cost of extra signal processing power consumption from the non-orthogonal ID signal detector. Therefore, the deployment of WDS is flexible and its computational complexity is related to applications and downlink/uplink channels.

In terms of hardware architectures, unlike \ac{MIMO} beamforming requiring multiple RF chains; millimeter wave requiring high frequency modulators; directional modulation requiring unique antennas; artificial noise requiring beamforming and power allocation, the proposed waveform-defined security framework can use available hardware and will not require additional hardware resources. Therefore, the hardware complexity of WDS is identical to the typical WLAN configurations. In this case, hardware complexity is not considered in this section.

The WDS computational complexity is thus analyzed merely on the digital signal generation. As explained in \cite{FFTW3_2005}, by using the framework of {\ac{FFTW}}, the asymptotic computational complexity of \ac{IDFT} is $\mathcal{O}(\xi\times{log_2{\xi}})$ when the transform size is $\xi$. It is well noticing that FFTW works well when the value of $\xi$ is either a power of two or a prime number. Therefore, considering the IDFT-based signal generation with $Q$ samples, the complexity of OFDM signal generation is
\begin{equation}\label{eq:OFDM_signal_gen_complexity}
\mathcal{O}(Q\times{log_2{Q}}).\end{equation}

In terms of SEFDM signal processing at the transmitter, as explained in Section \ref{sec:waveform_fundamental}, an IDFT architecture is applicable even when $\alpha$ is introduced. Therefore, the SEFDM signal generation complexity, computed based on \eqref{eq:signal_single_IFFT_pad_zeros} considering $Q/\alpha$ signal length, is given by
\begin{equation}\label{eq:SEFDM_signal_gen_complexity}
\mathcal{O}(Q/\alpha\times{log_2{Q/\alpha}}).\end{equation}

As explained in \eqref{eq:symbol_vector_single_IFFT_SEFDM}, zeros are padded at the end of each input symbol vector. Therefore, a pruned operation \cite{PaulVLSI} can be applied to simplify further the SEFDM signal generation complexity to
\begin{equation}\label{eq:SEFDM_signal_pruned_gen_complexity}
\mathcal{O}(Q/\alpha\times{log_2{Q}}).\end{equation}

\begin{figure}[t!]
\begin{center}
\includegraphics[scale=0.6]{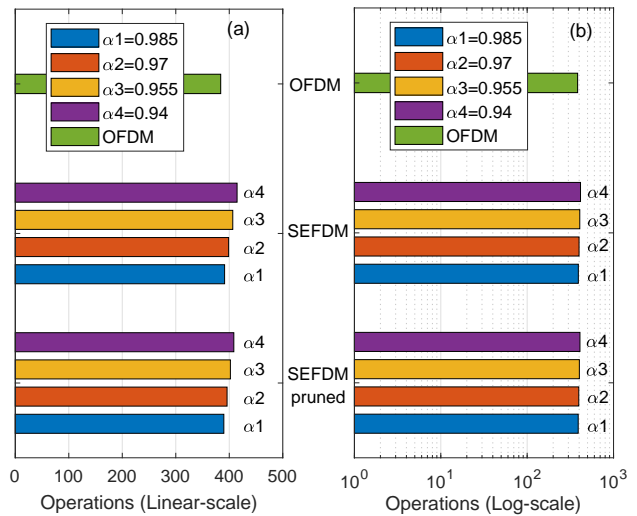}
\end{center}
\caption{{Asymptotic complexity of WLAN-Type-III framework in terms of the number of complex operations in (a) linear-scale and (b) log-scale.}}
\label{Fig:WDS_complexity}
\end{figure}

A bar chart is designed in Fig. \ref{Fig:WDS_complexity} to compare each signal generation method as a function of the number of operations. Computational complexity for each signal is computed following \eqref{eq:OFDM_signal_gen_complexity}, \eqref{eq:SEFDM_signal_gen_complexity} and \eqref{eq:SEFDM_signal_pruned_gen_complexity}. It is clearly seen in Fig. \ref{Fig:WDS_complexity} that the OFDM signal generation complexity is independent and will not be affected by the value of $\alpha$. The complexity of SEFDM signal generation is correlated to the value of $\alpha$. With the reduction of $\alpha$, more operations are required. This is reasonable since the value of $Q/\alpha$ is increased when $\alpha$ is reduced. Therefore, a larger IDFT is needed for an SEFDM signal with smaller $\alpha$. In addition, the pruned version of SEFDM signal generation has limited computational complexity reduction advantage. Another discovery in Fig. \ref{Fig:WDS_complexity} is that the number of required operations for SEFDM signal generation at different $\alpha$ is on the same order of magnitude relative to the OFDM signal generation.

\section{Low-Cost SDR Experiment} \label{sec:SDR_experiment}

Proof-of-concept experiments of traditional physical layer security techniques have been designed and tested in millimeter wave \cite{PLS_practical_CNS_mmWave_2015}, artificial noise generation \cite{PLS_practical_GC_AN_2017}, directional modulation \cite{PLS_practical_DM_2010} and secrecy coding \cite{PLS_practical_VTC_coding_2018}. However, those experiments might not be practical to low-cost and resource-constrained communication scenarios.

This section aims to verify the WDS framework in low-cost hardware with the following objectives:

\begin{itemize}

\item{ The WDS framework is applicable to low-cost hardware, which will be beneficial to resource-constrained IoT communications. } 

\item{ The WDS framework is highly coexistent with existing WLAN communication standards. It will be flexibly extended to other standards.} 

\item{ The WDS framework is robust and will not be compromised even when eavesdroppers have signal power and channel environment advantages. } 

\item{ CSIT is not required by the framework in experiments. } 

\end{itemize}

\subsection{Experiment Design}

\begin{figure}[t!]
\begin{center}
\includegraphics[scale=0.6]{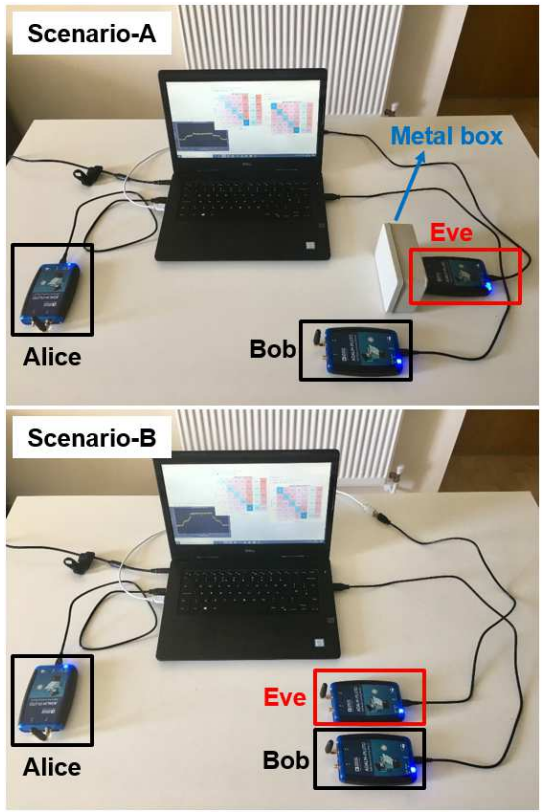}
\end{center}
\caption{Experiment setup for the WDS framework considering a non-line-of-sight eavesdropping link in Scenario-A and a line-of-sight eavesdropping link in Scenario-B.}
\label{Fig:WDS_WLAN_experiment_setup_Scenarios_A_B_pure_testbed}
\end{figure}

The low-cost Analog Devices \ac{SDR} PLUTO \cite{PlutoSDR} is applied in the experiment to demonstrate that the proposed WDS framework is practical to resource-constrained IoT scenarios. The framework does not require complex hardware such as \ac{MIMO} antennas, directional modulation driven antennas, phase shifters, high-frequency modulators and artificial noise generators. In this experiment, the SDR devices are equipped with single omni-directional antennas and therefore beamforming is not implemented.

The SDR device at Alice will generate all the signals based on the WLAN-Type-III signal pattern and deliver the signals over the air. Eve and Bob will receive them on their own SDR devices. Bob will demodulate and detect the signals with pre-shared signal format information while Eve will need to intelligently identify different signals before subsequent signal demodulation and detection.

To maintain a stable legitimate channel environment, the locations of Alice and Bob are fixed throughout the experiment. To have different eavesdropping channel scenarios, the position of Eve can be flexibly re-located. Two experiment scenarios are designed and tested using three SDR devices as shown in Fig. \ref{Fig:WDS_WLAN_experiment_setup_Scenarios_A_B_pure_testbed}.

In the first experiment, Scenario-A, a line-of-sight legitimate link exists between Alice and Bob while the eavesdropping link between Alice and Eve is blocked by a metal box to emulate a non-line-of-sight communication. This scenario is challenging to Eve since the eavesdropping link signal power is lower than that of legitimate link. This scenario is designed based on the assumption that Eve in this work is passive and will merely listen to confidential information in a disadvantageous location. Otherwise, the existence of Eve will be detected.

In the second experiment, Scenario-B, Eve is placed next to Bob without the metal box blockage. In this case, a line-of-sight eavesdropping link is created, which has a similar signal power condition with the legitimate link. Scenario-B is challenging to communication security since the eavesdropper has a better condition than that in Scenario-A. Typical beamforming and artificial noise based physical layer security solutions are not efficient any more since Eve and Bob are placed closely next to each other.

This experiment will follow the 802.11a signal specifications and use the maximum bandwidth option, 20 MHz. The employed SDR PLUTO is supported by the \ac{WLAN} toolbox \cite{WLAN_matlab_toolbox} in Matlab. Therefore, the implementations of 802.11a and WLAN-WDS are straightforward. Over-the-air carrier frequency is tuned to 2.412 GHz, which is the Channel-1 frequency defined by \cite{IEEE_802_11}. To have a high coexistence with WLAN, this experiment maintains the 802.11a standard defined legacy preamble while merely changing the PSDU data field to the WLAN-Type-III signal pattern.

\subsection{Experiment Results}

To validate the experimental communication security at the eavesdropper, both confusion matrix and BER are investigated. In the experiment, 1,000 symbols per signal class are generated. There are overall 5,000 OFDM/SEFDM symbols for each test. Unlike simulations, in practical experiments, \ac{SNR} is commonly measured at the receiver. In this experiment, to show the signal power difference between Eve and Bob, \ac{SNR} is measured based on their frequency-domain signal spectra.

\begin{figure}[t!]
\begin{center}
\includegraphics[scale=0.8]{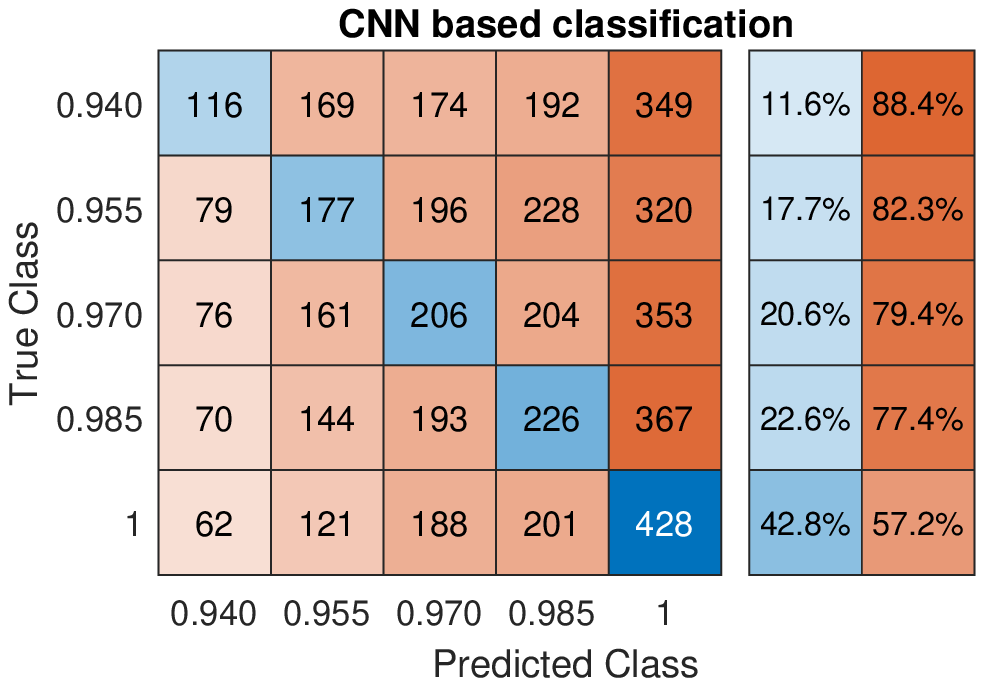}
\end{center}
\caption{Confusion matrix of Eve employing CNN classification in Scenario-A. $\alpha$=1 indicates WLAN-OFDM while other values of $\alpha$ indicate WLAN-SEFDM.}
\label{Fig:WDS_confusion_matrix_scenario_A_CNN}
\end{figure}

\begin{figure}[t!]
\begin{center}
\includegraphics[scale=0.8]{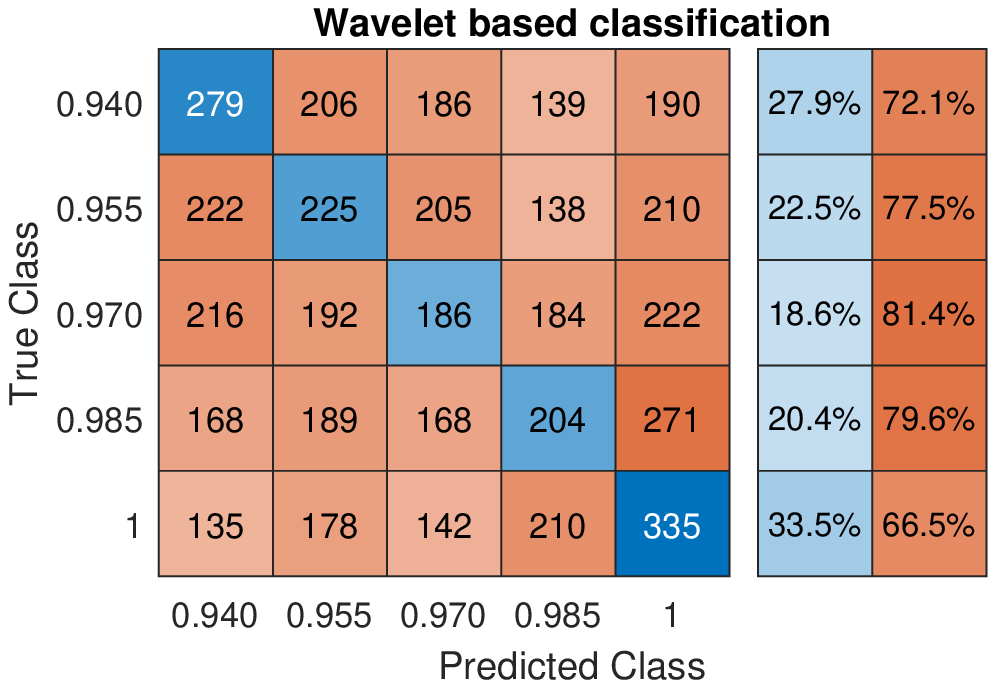}
\end{center}
\caption{Confusion matrix of Eve employing wavelet classification in Scenario-A. $\alpha$=1 indicates WLAN-OFDM while other values of $\alpha$ indicate WLAN-SEFDM.}
\label{Fig:WDS_confusion_matrix_scenario_A_wavelet}
\end{figure}

In Scenario-A, the value of \ac{SNR} at Bob is around 35 dB, which is power sufficient to provide reliable performance with zero BER. The metal box at Eve can create a non-line-of-sight link but cannot influence SNR greatly. Therefore, by tuning the \ac{AGC} at Eve, the SNR is reduced to 10 dB. This will emulate the practical eavesdropping condition where interception environment is commonly disadvantageous. With the above experiment setup, the average successful classification accuracy at Eve is 23.06\% for the CNN classifier in Fig. \ref{Fig:WDS_confusion_matrix_scenario_A_CNN} and 24.58\% for the wavelet classifier in Fig. \ref{Fig:WDS_confusion_matrix_scenario_A_wavelet}.

\begin{figure}[t!]
\begin{center}
\includegraphics[scale=0.8]{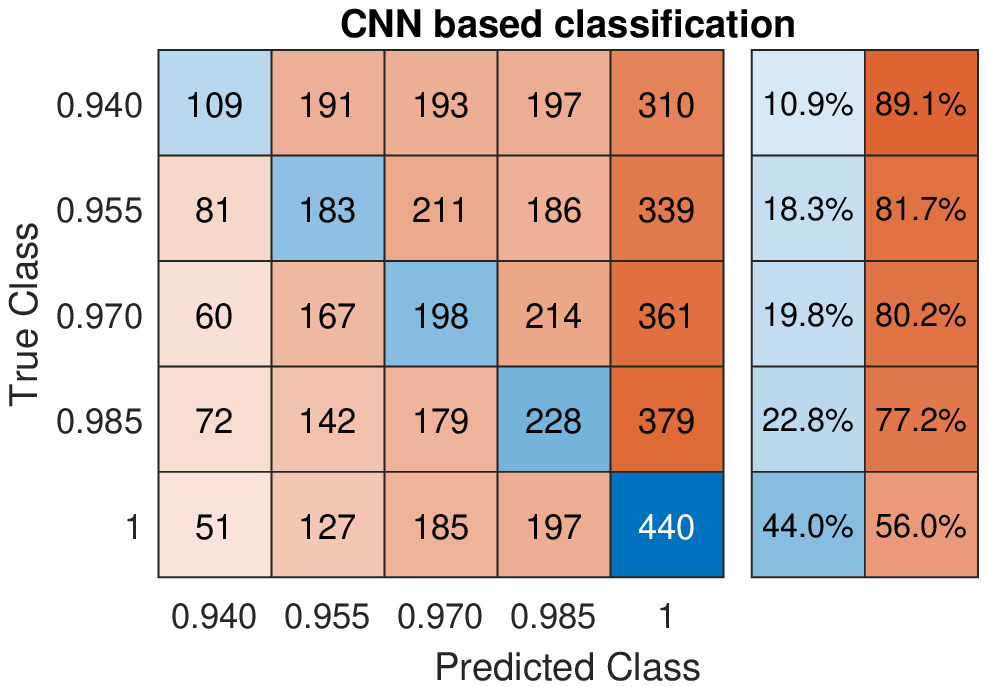}
\end{center}
\caption{Confusion matrix of Eve employing CNN classification in Scenario-B. $\alpha$=1 indicates WLAN-OFDM while other values of $\alpha$ indicate WLAN-SEFDM.}
\label{Fig:WDS_confusion_matrix_scenario_B_CNN}
\end{figure}

\begin{figure}[t!]
\begin{center}
\includegraphics[scale=0.8]{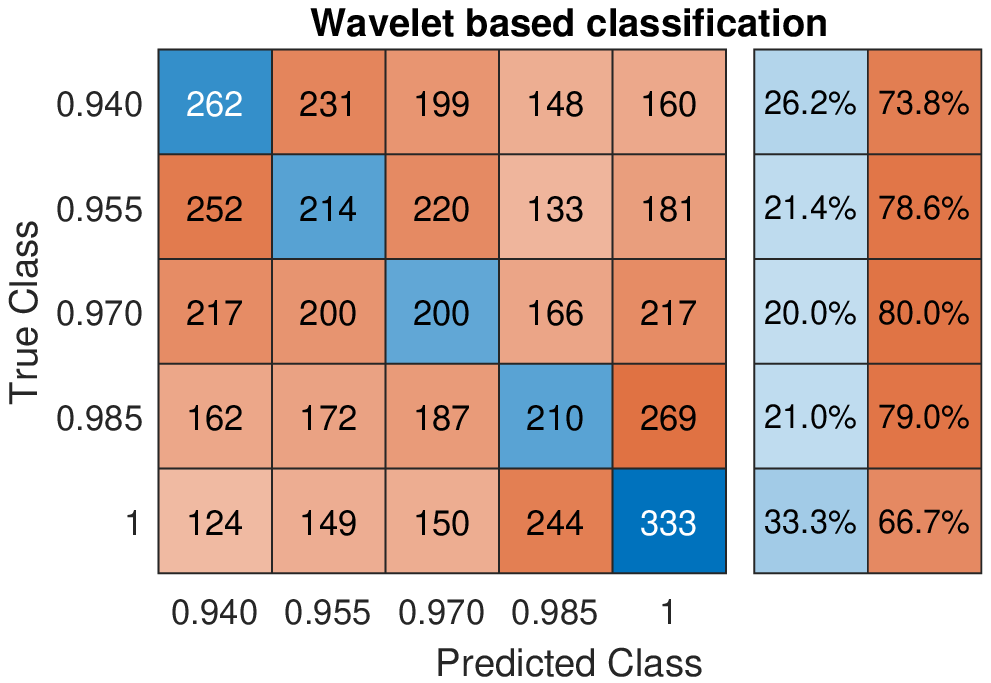}
\end{center}
\caption{Confusion matrix of Eve employing wavelet classification in Scenario-B. $\alpha$=1 indicates WLAN-OFDM while other values of $\alpha$ indicate WLAN-SEFDM.}
\label{Fig:WDS_confusion_matrix_scenario_B_wavelet}
\end{figure}

In Scenario-B, the location of Bob is fixed while Eve is placed next to Bob, which gives Eve a better eavesdropping environment with a line-of-sight link at SNR=35 dB. The confusion matrices for Scenario-B are presented in Fig. \ref{Fig:WDS_confusion_matrix_scenario_B_CNN} and Fig. \ref{Fig:WDS_confusion_matrix_scenario_B_wavelet}, which show average classification accuracy of 23.16\% and 24.38\%, respectively. It is practically verified that the proposed waveform-defined security framework is robust and is not related to SNR and communication link conditions at eavesdroppers.

\begin{figure}[t!]
\begin{center}
\includegraphics[scale=0.58]{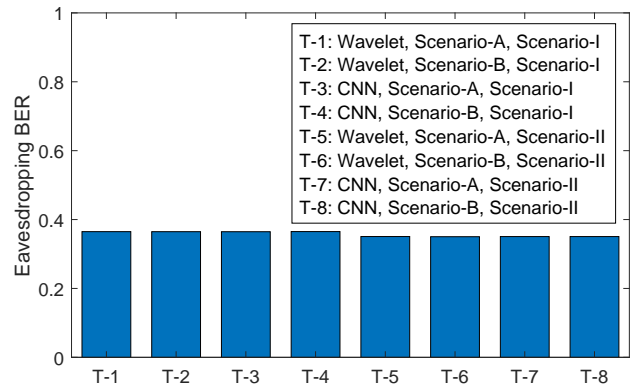}
\end{center}
\caption{Eavesdropping BER considering signal classifiers, eavesdropper channel conditions and eavesdropper receiver operations. }
\label{Fig:WDS_BER_eavesdropping_mixed_scenarios}
\end{figure}

An additional experiment discovery is that the CNN classification accuracy is not equal for each signal class in Fig. \ref{Fig:WDS_confusion_matrix_scenario_A_CNN} and Fig. \ref{Fig:WDS_confusion_matrix_scenario_B_CNN}. It is apparent that WLAN-OFDM has a higher classification accuracy than other signal types achieving 42.8\% and 44.0\% in Scenario-A and Scenario-B, respectively. Moreover, all other WLAN-SEFDM signals are mostly classified into WLAN-OFDM. The reason for this is that the offline trained CNN model is not perfectly fit in a new channel environment. Previous work \cite{tongyang_VTC2020_DL_classification} applied transfer learning to deal with the mismatch between offline training environment and practical environment. However, this is not realistic in secure communications since legitimate users will not allow eavesdroppers to adjust signal classifiers via transfer learning. Therefore, the mismatch between offline and practical environments would additionally enhance communication security. In terms of wavelet classification, accuracy is relatively stable for each signal class, which indicates the robustness of wavelet classification since signal features are manually extracted based on domain knowledge rather than data training.

To have a comprehensive study on eavesdropping BER performance, Fig. \ref{Fig:WDS_BER_eavesdropping_mixed_scenarios} jointly considers signal classifier types (i.e. CNN and wavelet), eavesdropper channel conditions (i.e. Scenario-A and Scenario-B) and eavesdropper receiver operations (i.e. Scenario-I and Scenario-II). Therefore, eight tests are designed with results showing in Fig. \ref{Fig:WDS_BER_eavesdropping_mixed_scenarios}. It is clear that all the BER results maintain at a similar level while the Scenario-II based eavesdropping shows slightly degraded performance. As explained in Section \ref{subsec:reliability_security}, Scenario-I has no classification mechanism and assumes all the received signals belong to WLAN-OFDM. However, Scenario-II applies classifiers and misclassification would happen. The performance of signal detection is highly dependent on the quality of signal classification. It is inferred from Fig. \ref{Fig:WDS_BER_eavesdropping_mixed_scenarios} that signal misclassification might leads to better BER than a system without signal classification.

\section{Conclusion} \label{sec:conclusion}

Traditional physical layer security techniques are highly dependent on channel conditions such as accurate CSI at the transmitter. However, in practical communications, perfect CSI is mostly unachievable due to time-variant channel characteristics. In addition, extra hardware is commonly required by beamforming or artificial noise techniques. Therefore, a waveform-defined security (WDS) framework is proposed to avoid the dependance on CSI and complex hardware. This work firstly studies three impact factors, which can tune waveform patterns undiscoverable at eavesdroppers. Results show that training data diversity has great effects on signal identification accuracy. In addition, a small oversampling factor and a narrow BCF offset can further confuse eavesdropping. A WLAN-WDS frame is designed to show a compatible coexistence with the existing WLAN 802.11a standard. Results show that BER performance is ensured at the legitimate user and security is promised via preventing eavesdropping. In addition, WLAN-WDS has the computational complexity on the same order of magnitude relative to the traditional WLAN. A low-cost experiment is operated to verify the WDS framework in resource-constrained communication systems. The SDR devices applied in this work are equipped with single omni-directional antennas and therefore beamforming is not implementable. Results show that the eavesdropper fails to recover signals even with an advantageous line-of-sight channel condition at SNR=35 dB, which proves that the success of WDS framework is not dependent on channel conditions. This work successfully verified that the proposed WDS framework is applicable in resource-constrained IoT communications while traditional PLS techniques are unachievable due to channel condition dependance and extra hardware complexity.

\bibliographystyle{IEEEtran}
\bibliography{WDS_JIoT_Ref}

\end{document}